\documentclass[aps,prl,twocolumn,a4paper,showpacs,superscriptaddress]{revtex4-1}

\usepackage{graphicx,xcolor}
\usepackage{dcolumn}
\usepackage{bm}

\usepackage{hyperref}

\begin{document}

\title{Network of topological nodal planes, multifold degeneracies, and Weyl points in CoSi}

\author{Nico Huber}
\thanks{These authors contributed equally.}
\affiliation{Physik Department, Technische Universit\"at M\"unchen, D-85748 Garching, Germany}
\author{Kirill Alpin}
\thanks{These authors contributed equally.}
\affiliation{Max-Planck-Institute for Solid State Research, Heisenbergstrasse 1, D-70569 Stuttgart, Germany}
\author{Grace L. Causer}%
\affiliation{Physik Department, Technische Universit\"at M\"unchen, D-85748 Garching, Germany}
\author{Lukas Worch}%
\affiliation{Physik Department, Technische Universit\"at M\"unchen, D-85748 Garching, Germany}
\author{Andreas Bauer}%
\affiliation{Physik Department, Technische Universit\"at M\"unchen, D-85748 Garching, Germany}
\author{Georg Benka}%
\affiliation{Physik Department, Technische Universit\"at M\"unchen, D-85748 Garching, Germany}

\author{Moritz~M. Hirschmann}
\affiliation{Max-Planck-Institute for Solid State Research, Heisenbergstrasse 1, D-70569 Stuttgart, Germany}
\author{Andreas P. Schnyder}
\affiliation{Max-Planck-Institute for Solid State Research, Heisenbergstrasse 1, D-70569 Stuttgart, Germany}
\email{a.schnyder@fkf.mpg.de}

\author{Christian Pfleiderer}%
\affiliation{Physik Department, Technische Universit\"at M\"unchen, D-85748 Garching, Germany}
\affiliation{MCQST, Technische Universit\"at M\"unchen, D-85748 Garching, Germany}
\affiliation{Centre for Quantum Engineering (ZQE), Technische Universit\"at M\"unchen, D-85748 Garching, Germany}
\author{Marc A. Wilde}
\email{mwilde@ph.tum.de}
\affiliation{Physik Department, Technische Universit\"at M\"unchen, D-85748 Garching, Germany}

\date{\today}

\begin{abstract}
We report the identification of symmetry-enforced nodal planes (NPs) in CoSi providing the missing topological charges in an entire network of band-crossings comprising in addition multifold degeneracies and Weyl points, such that the fermion doubling theorem is satisfied. In our study we have combined measurements of Shubnikov--de Haas (SdH) oscillations in CoSi with material-specific calculations of the electronic structure and Berry curvature, as well as a general analysis of the band topology of space group (SG)\,198. The observation of two nearly dispersionless SdH frequency branches provides unambiguous evidence of four Fermi surface sheets at the R point that reflect the symmetry-enforced orthogonality of the underlying wave functions at the intersections with the NPs. Hence, irrespective of the spin-orbit coupling strength, SG\,198 features always six- and fourfold degenerate crossings at R and $\Gamma$ that are intimately connected to the topological charges distributed across the network.
\end{abstract}

\maketitle

In recent years the identification of non-trivial topological quasiparticles (TQPs) in solids attracts great interest. Contributions of such topological quasiparticles in the physical properties are expected to scale with the Berry curvature of the bands, the inverse energy difference between the band crossings and the Fermi level, and the separation of corresponding sources and sinks of Berry curvature in $k$-space. Maximally large Chern numbers in nonmagnetic chiral materials crystallizing in SG 198, notably CoSi, RhSi, and PdGa \cite{CoSi_PRL_17, ChangHasan-RhSi-PRL-2017, Pshenay_Severin_2018, CoSi_Nature_hong_19, RhSi_Nature_hasan_19, 2019-Takane-PRL}, were recently inferred from the observation of Fermi arcs \cite{CoSi_Nature_hong_19, RhSi_Nature_hasan_19, 2019-Takane-PRL, 2020_Schroeter_Science} and quasiparticle interference patterns \cite{2019-Yuan-Science-Advances}. Based on the bulk-boundary correspondence and ab-initio calculations these observations were attributed to degeneracies at the $\Gamma$ and R point and corresponding charges believed to satisfy the fermion doubling theorem (FDT) between them~\cite{supp,nielsen_no_go}. However, the discovery of symmetry-enforced topological nodal planes (NPs) in the ferromagnetic state of MnSi \cite{MnSi_nodal_plane}, described by magnetic subgroups of SG\,198, raises the questions for (i) the existence of topological NPs in SG\,198, (ii) whether the analysis of the band topology and the Berry curvature for specific points only is sufficient for an assessment of their impact on the physical properties, and, (iii) whether strongly simplified Hamiltonians of the TQPs as inspired by high-energy physics \cite{CoSi_PRL_17} are justified \cite{Pshenay_Severin_2018}.  

When spin-orbit coupling (SOC) is strong as in PdGa, spin degeneracies are lifted, leading to fourfold and twofold points at $\Gamma$ and a sixfold point at R \cite{2020_Schroeter_Science}. In contrast, it has been argued that SOC may be ignored in the light transition-metal silicides CoSi and RhSi such that all bands are spin-degenerate, leading to a threefold point at $\Gamma$ and a fourfold point at R \cite{Chern_count}. In fact, scanning tunneling microscopy studies in CoSi suggest a SOC as large as 25 to 35\,meV \cite{2019-Yuan-Science-Advances}, while conflicting interpretations of quantum oscillations in CoSi, inferred a SOC as low as one meV \cite{Xu_2019_PRB, Wu_SdH-CPL-2019, Wang-dHvA-PRB-2020}. This raises the question of how to reconcile the topological properties of SG\,198 with SOC and NPs, being dictated by fundamental quantum theory and crystal symmetry.

\begin{figure}[b]
 \includegraphics[width=8.6cm]{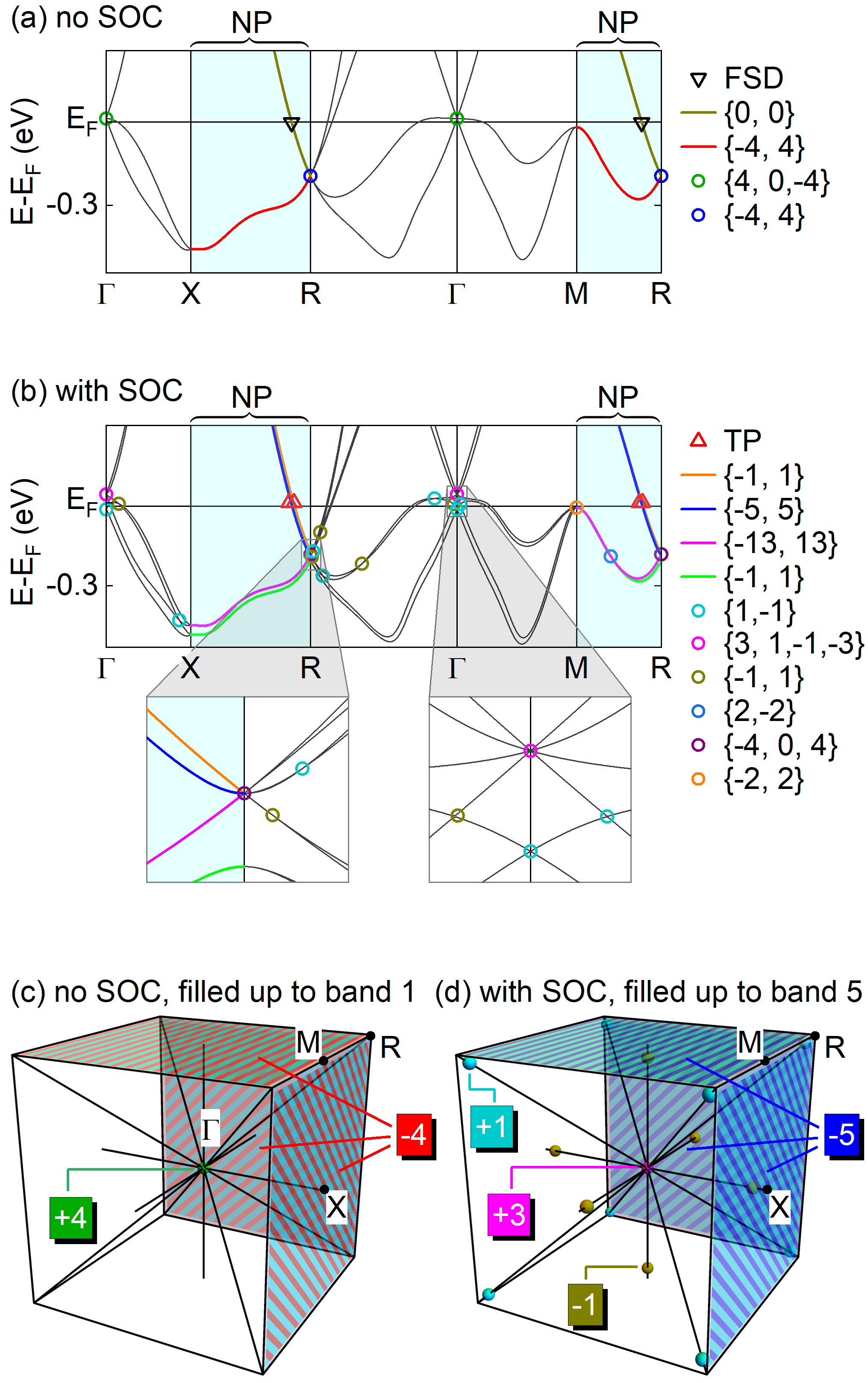}
	\caption{\label{fig:bands-and-FS} Electronic band structure of CoSi. See text for details of notation.  (a) Band structure without SOC.  (b) Band structure with SOC. Insets: close-up views around R and $\Gamma$. (c, d) Illustration of the FDT for selected band fillings. No SOC and filling up to band 1 yields a charge of $+4$ at $\Gamma$ (green) compensated by the trio of NPs with charge $-4$ (red/cyan). With SOC and filling up to band 5 the bulk crossing points carry a net charge of $+5$, taking into account their multiplicities, which is compensated by the trio of NPs with charge $-5$ (blue/cyan). }
\end{figure}

As SG~198 is chiral with three two-fold screw rotations along the three main axes, any band structure in this SG contains NPs on the three Brillouin zone (BZ) boundaries $k_{x,y,z} = \pm \pi$.  This trio of NPs can carry a finite topological charge that is protected by the screw rotations combined with time-reversal symmetry~\cite{MnSi_nodal_plane,PRB_Yuxin,2019_PRB_Yuxin}. 

Inferring the existence of band crossings on high-symmetry points and lines from the dimensionality of the irreps of the little groups and their compatibility relations, we performed in this work a general band topology analysis and determined the main aspects of NPs in SG\,198 (cf. Supplemental Material (SM)~\cite{supp} for details).
For instance, with SOC there can be twofold and fourfold degeneracies at $\Gamma$, while on the $\Gamma$--X-path there are symmetry-enforced, movable Weyl crossings. With SOC the pinned crossings at $\Gamma$ enforce the topological charge of the NPs. Finally, the multiplicity of accidental crossing points at and away from high-symmetry lines or planes is independent of SOC, and the Chern number of symmetry-related Weyl points is always the same since SG\,198 does not contain inversion or mirror symmetries. Thus, a symmetry-enforced topological charge of the NPs must be \emph{odd}.

To confirm these general findings we studied CoSi, calculating the band structure and Fermi surface (FS) in density functional theory (DFT) using \textsc{WIEN2k} \cite{WIEN2k} and \textsc{Quantum Espresso} \cite{giannozzi2009quantum,DALCORSO2014337} within the generalized gradient approximation \cite{PBE_functional}. We then determined the band topology and the topological charges of the trio of NPs, computing the Berry curvature from the DFT wave functions \cite{fukui_JPSJ_05, wannier90, pizzi2020wannier90, gosalbez2015chiral} (for details see SM~\cite{supp}).

In the following we denote the Chern number, $\nu$, and the multiplicity, $l$, of band crossings with and without SOC by $\nu^n_A$,  $l^n_A$  and $\tilde{\nu}^n_A$, $\tilde{l}^n_A$, respectively \cite{supp}. Here $A$ denotes the location in the BZ and $n$ designates the index of the energetically lower band (or bandpair when the crossing is located on a NP) that participates in the degeneracy going from low to high energies starting above the global band gap at $\sim -0.5$\,eV.  While we do not denote the spin degeneracy explicitly when labeling bands, we include it in the Chern numbers.

Summarized in Fig.~\ref{fig:bands-and-FS} are the band structures of CoSi without and with SOC as well as illustrations of key results of our band topology analysis. Cyan shading highlights the NPs. Bands intersecting with the NPs are marked in color. Topological point crossings are marked by colored circles. Curly brackets denote all well-defined Chern numbers of the crossing bands in ascending order in energy. Fermi surface degeneracies (FSDs) and topological protectorates (TPs) \cite{MnSi_nodal_plane}, i.e., topologically non-trivial FSDs, are marked by triangles.

Consistent with the underlying symmetries, the band structure of CoSi exhibits NPs on the three BZ boundaries regardless of SOC. Without SOC all bands are spin-degenerate [Fig.~\ref{fig:bands-and-FS}(a)], and there are two spin-degenerate trios of NPs close to the Fermi energy $E_{\rm F}$, where the topological charges for each of the crossings add to zero in accordance with the FDT. With SOC the spin degeneracy is lifted [Fig.~\ref{fig:bands-and-FS}(b)] and the two trios of NPs split into four and the corresponding crossings at $\Gamma$ are forced to carry odd topological charges \cite{CoSi_PRL_17, Pshenay_Severin_2018}. For these odd band fillings $n$ the charges at $\Gamma$ can only be compensated by the trios of NPs, as all other crossings (accidental or enforced) sum to an even number~\cite{supp,MnSi_nodal_plane}. Thus, with SOC the two NP trios crossing $E_{\rm F}$ form TPs. 

To determine the relationship between the different topological charges, we analyzed the band structure of CoSi systematically for different band fillings. Without SOC we found that the NPs are not forced by symmetry to carry a charge. In contrast, with SOC all NPs are forced to carry charges by symmetry. 

Two selected band fillings nicely illustrate this situation (for all other fillings see SM\,\cite{supp}). First, without SOC and the first band being filled, the trio of NPs that is lower in energy, marked in red/cyan in Fig.~\ref{fig:bands-and-FS}(c), supports a charge of $\nu^1_{\rm npt}=-4$ that compensates the charge of $\nu^1_{\rm \Gamma}=+4$ of the threefold point at $\Gamma$ (green), in agreement with the FDT~\cite{nielsen_no_go}. The upper trio of NPs, in contrast, has charge zero, since the Chern numbers at $\Gamma$ given by $(\nu^1_{\mathrm{\Gamma}},\nu^2_{\mathrm{\Gamma}},\nu^3_{\mathrm{\Gamma}}) = \{4,0,-4\}$ add up to zero [green circle in Fig. 1(a)], when the lowest three bands are filled. 
Second, when taking into account SOC, the situation changes fundamentally as illustrated in Fig.~\ref{fig:bands-and-FS}(d) for a filling up to band 5. Here we find at $\Gamma$ a total charge $\nu_{\Gamma}^{3} + \nu_{\Gamma}^{4} + \nu_{\Gamma}^{5} = +3$. On the $\Gamma$--X line there is a movable Weyl point with charge $\nu^5_{\Gamma\text{X}} =-1$ and multiplicity $l^5_{\Gamma\text{X}} =6$, resulting in a charge  $l^5_{\Gamma\text{X}}\,\nu^5_{\Gamma\text{X}} = -6$. On the $\Gamma$--R line there is a crossing, $\nu^5_{\Gamma\text{R}} = +1$ very close to R, contributing a total charge $l^5_{\Gamma\text{R}}\,\nu^5_{\Gamma\text{R}} = 8$. Since an odd number of bands is filled, the point crossings at M and R within the NP do not contribute any charge. Hence, according to the FDT the NP must compensate these charges, i.e.,  $  \nu^5_{\mathrm{npt}} = - \nu_{\Gamma}^{3} - \nu_{\Gamma}^{4} - \nu_{\Gamma}^{5} - 6\nu^5_{\Gamma\text{X}} - 8\nu^5_{\Gamma\text{R}}  = - 3 +6 -8 = -5$. This result is also obtained in direct numerical calculations \cite{supp}, indicating the absence of additional uncompensated Weyl points at generic positions.

\begin{figure}[b]
	\includegraphics[width=8.6cm]{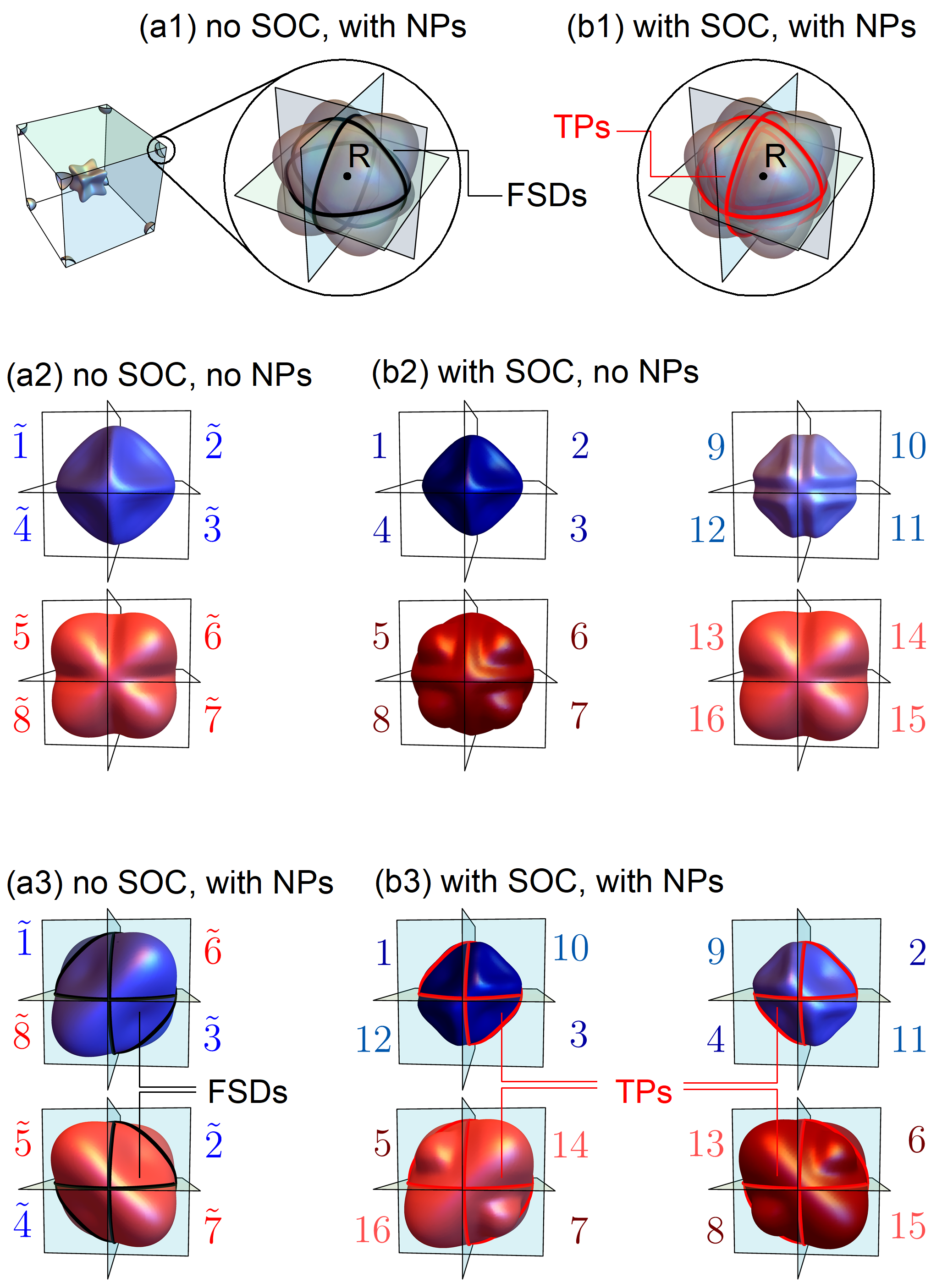}%
	\caption{\label{fig:fs-sketches} Effect of SOC and NPs on the FS sheets at the R point. See text for details of notation. (a1, b1) FS sheets at the R point and splitting without and with SOC.  (a2, a3) FS sheets at the R point without SOC. Two spin-degenerate sheets are expected. NPs alter the band connectivity such that diagonally located octants combine to form the FS sheets. 
	(b2, b3) FS sheets at the R point with SOC. Effect of the NPs is the same as in (a3).}
\end{figure}

To prove the existence of NPs experimentally, we measured the quantum oscillations arising from the FS sheets centred at the R point as these sheets intersect with the NPs (Fig.\,\ref{fig:fs-sketches}) \cite{Xu_2019_PRB, Wu_SdH-CPL-2019, Wang-dHvA-PRB-2020}. Here, FSDs and TPs denote the location of FS degeneracies (black line) and topological protectorates (red line), respectively and the numbers denote the octants of the FS sheets facing the observer. A constant length has been subtracted from the distance between R and the FS contour for better visibility of the different shapes. The predicted SOC splitting at $E_{\rm F}$ around R is small $\sim 15$\,meV, for low-symmetry directions. This corresponds to a $k$-space distance of the FS sheets of about $\sim 0.7$\,\% of a reciprocal lattice vector, which may be readily resolved in quantum oscillations.

Without SOC two spin-degenerate sheets are expected [Fig.\,\ref{fig:fs-sketches}\,(a1) to \ref{fig:fs-sketches}\,(a3)], which exhibit topologically trivial FSDs on the NPs [Fig.\,\ref{fig:fs-sketches}\,(a1)]. Ignoring the NPs [Fig.\,\ref{fig:fs-sketches}\,(a2)], the extremal cross-sectional areas are identical for $B\parallel [100]$, resulting in a single oscillation frequency [cf. Fig.\,\ref{fig:R-point-explanation}\,(a)]. In comparison, for arbitrary directions the FS cross-sections are different [Fig.\,\ref{fig:fs-sketches}\,(a2)] and two dispersive frequency branches are expected [cf. Fig.\,\ref{fig:R-point-explanation}\,(a)]. Considering now the NPs [Fig.\,\ref{fig:fs-sketches}\,(a3)], the band connectivity changes such that (space-) diagonally located octants are recombined, i.e., sectors $\tilde{1}$ and $\tilde{3}$ in Fig.\,\ref{fig:fs-sketches}\,(a2) combine with $\tilde{6}$ and $\tilde{8}$ in Fig.\,\ref{fig:fs-sketches}\,(a2) (and analogously for the octants facing away from the observer) to form the blue FS sheet shown in Fig.\,\ref{fig:fs-sketches}\,(a3). Vice versa sectors $\tilde{2}$, $\tilde{4}$, $\tilde{5}$, and $\tilde{7}$ form the red FS sheet shown in Fig.\,\ref{fig:fs-sketches}\,(a3). The extremal cross-sections of the reconstructed FS sheets are essentially identical regardless of orientation, resulting in a single almost dispersionless frequency branch even though two FS sheets contribute [cf. Fig.\,\ref{fig:R-point-explanation}\,(c)]. Further, the band connectivity suppresses transitions between different sheets and thus additional branches. 

With SOC the spin degeneracy is lifted resulting in four FS sheets [Fig.\,\ref{fig:fs-sketches}\,(b1) to \ref{fig:fs-sketches}\,(b3)]. The analysis of the band topology reported above establishes TPs (red lines). The effects of the NPs are identical to the FS sheets without SOC. Due to the band connectivities at the NPs the four FS sheets reduce to two pairs with essentially identical cross-sections regardless of orientation [Fig.\,\ref{fig:fs-sketches}\,(b3)]. Thus, two frequency branches with low dispersion are expected even though four FS sheets contribute [cf. Fig.\,\ref{fig:R-point-explanation}\,(d)] This picture gets modified when additionally taking into account magnetic breakdown, which further decreases the angular dispersion of the two branches with the highest weight.

\begin{figure}[b]
	\includegraphics[width=8.6cm]{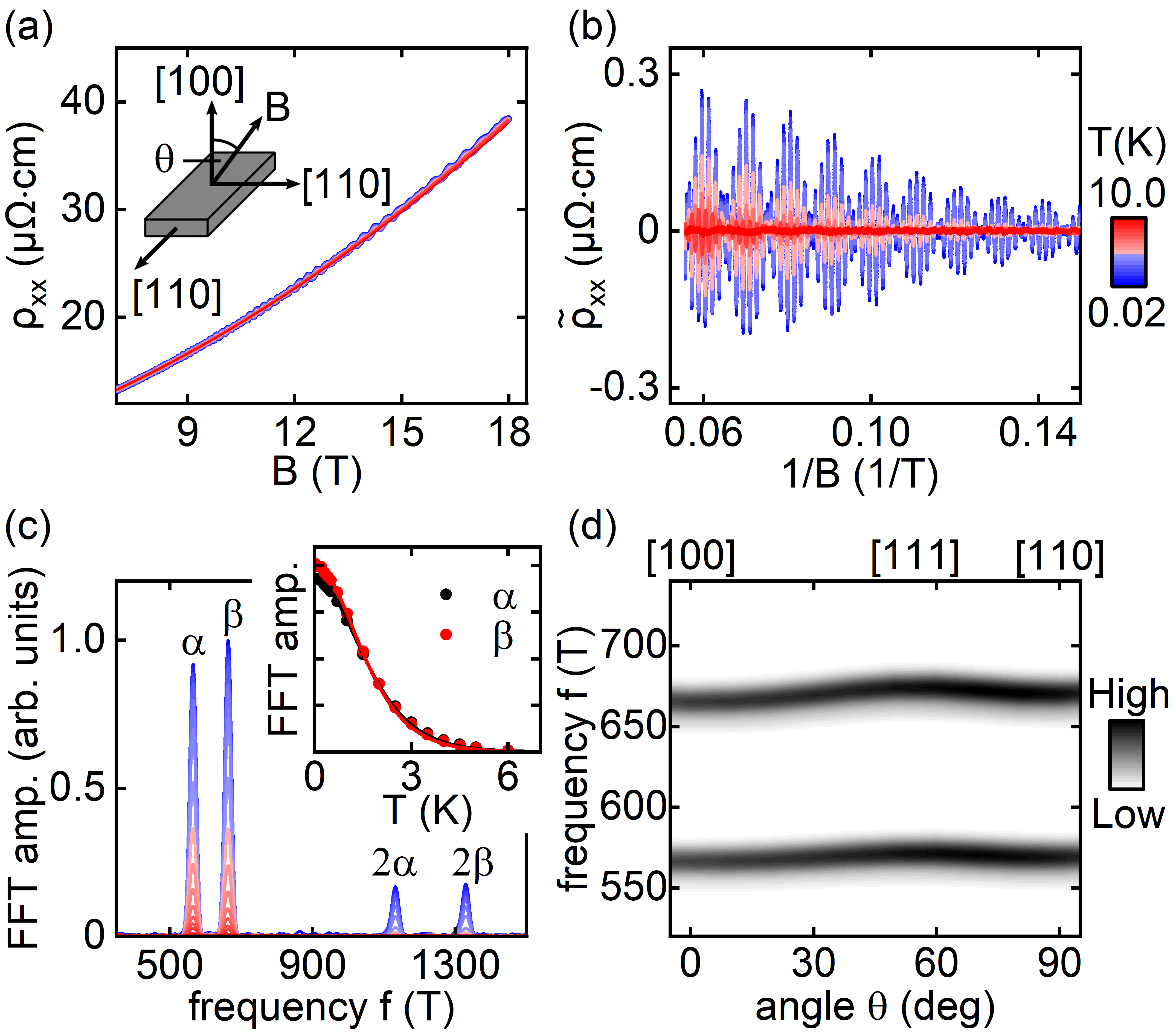}%
	\caption{\label{fig:experiment} Typical SdH data. (a) Magnetoresistance $\rho_{\text{xx}}(B)$; Inset: experimental geometry using cubic equivalent directions. (b) Oscillatory part of the resistivity $\tilde{\rho}_{xx}(1/B)$ for selected temperatures. (c) FFTs of $\tilde{\rho}_{xx}(1/B)$. Two frequencies $f_{\alpha}=565$\,T and $f_{\beta}=663$\,T and their harmonics are resolved. Inset: Lifshitz-Kosevich-behavior of the FFT amplitudes, yielding effective masses $m_{\alpha}=0.90$\,$m_e$ and $m_{\beta}=0.95$\,$m_e$. (d) FFT amplitude versus frequency and field direction $\theta$. Both SdH branches exhibit a very low dispersion. }
\end{figure}

To confirm our predictions for the FS sheets centered at R we measured the Shubnikov--de Haas (SdH) effect, where our data are in excellent agreement with the literature \cite{Xu_2019_PRB, Wu_SdH-CPL-2019, Wang-dHvA-PRB-2020}. For details on the experimental methods see SM \cite{supp}. As illustrated in Fig.~\ref{fig:experiment}\,(a), the resistivity $\rho_{\rm xx}$ exhibits pronounced quantum oscillations for magnetic fields $B \parallel [100]$ exceeding $\sim6\,{\rm T}$ on a monotonically increasing background, where the oscillatory component of the signal is shown in Fig.~\ref{fig:experiment}\,(b). Fast Fourier transforms (FFTs) were calculated using a Hamming window, where typical FFT spectra are shown in Fig.~\ref{fig:experiment}\,(c). 

Two frequencies, $f_{\alpha}=565$\,T and $f_{\beta}=663$\,T, and their higher harmonics could be resolved. Based on close inspection of the higher harmonics observed in samples of different quality, we conclude that asymmetries of the FFT peaks reported previously are not intrinsic (see SM\,\cite{supp} for details). The temperature dependence of the FFT amplitude was analyzed within the Lifshitz-Kosevich formalism \cite{Shoenberg1984}, yielding effective masses $m_{\alpha}=0.90\,m_e$ and $m_{\beta}=0.95\,m_e$ [inset of Fig.~\ref{fig:experiment}\,(c)]. The angular dispersion of the SdH branches is shown in Fig.~\ref{fig:experiment}\,(d), where the strength of the shading reflects the FFT amplitude. The variation of frequency with $\theta$ was very low around $\sim 5\,{\rm T}$ and $\sim 10\,{\rm T}$ for $f_{\alpha}$ and $f_{\beta}$, respectively. 

\begin{figure}[b]
	\includegraphics[width=8.6cm]{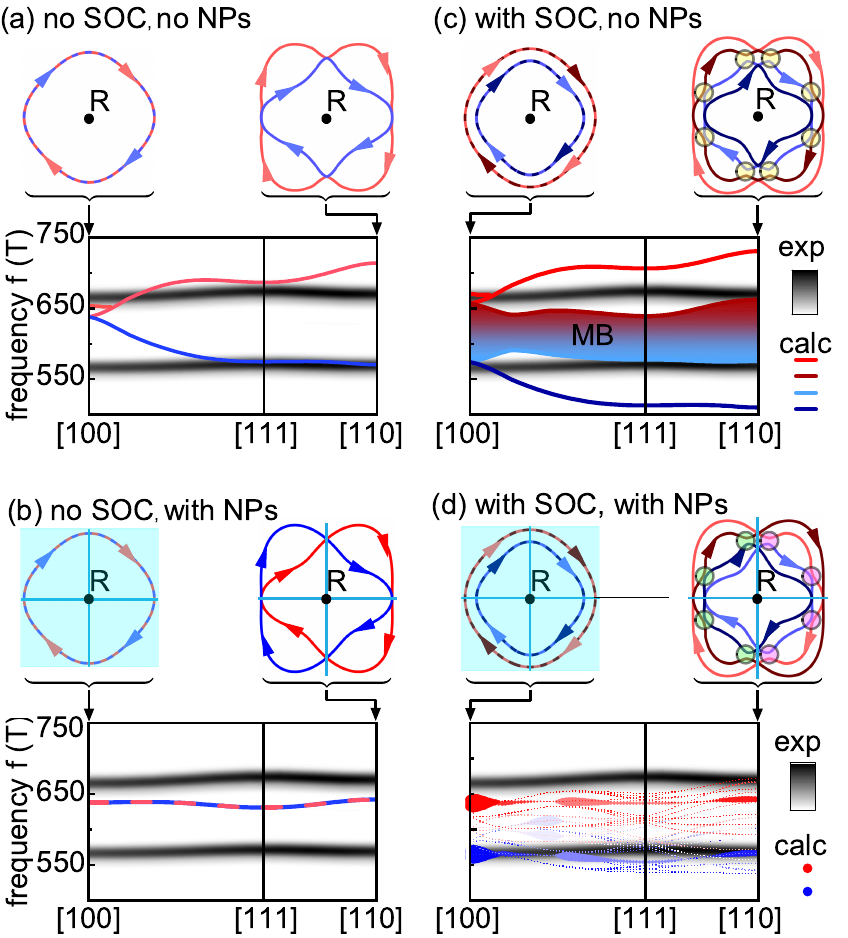}%
	\caption{\label{fig:R-point-explanation} Comparison of experimental SdH branches (gray scale) and calculated branches (color). From the shapes of extremal orbits for $B\parallel [100]$ and $B\parallel [110]$ an R-centered circle was subtracted for better visibility. Magnetic breakdown junctions are marked by colored circles. (a) Without SOC and without NPs. (b) With NPs (cyan) and without SOC. (c) With SOC and without NPs. Shaded area reflects up to 8192 partly degenerate breakdown branches. (d) With SOC and with NPs. Breakdown branches are calculated explicitly, resulting in two strong almost dispersionless branches.}
\end{figure}

For the interpretation of our data we assume a semiclassical quasiparticle motion and magnetic breakdown according to Chambers formula \cite{Shoenberg1984,supp}. Shown in Fig.~\ref{fig:R-point-explanation} are the cyclotron orbits for selected field orientations and comparisons of the experimental frequency branches (gray scale) with the calculated behaviors (colored lines) [Fig.\,\ref{fig:experiment}(a-d)].

Without SOC and NPs, representing the interpretation in Refs.\,\onlinecite{Wu_SdH-CPL-2019,Xu_2019_PRB}, there are two spin-degenerate bands crossing $E_{\rm F}$ near R. For $B\parallel [100]$ the two extremal orbits coincide [Fig.~\ref{fig:R-point-explanation}\,(a)], resulting in a single SdH frequency. A tiny splitting of the upper branch close to [100], reflecting the constrictions of the red FS sheet in Fig.~\ref{fig:fs-sketches}\,(a), vanishes quickly under field rotation, whereas the main branch splits into a dispersive inner and outer orbit in stark contrast with experiment.

Further, neglecting SOC but taking into account the band connectivity due the NPs [Fig.~\ref{fig:R-point-explanation}\,(b),] the extremal cross-sections of the two FS sheets will be the same, resulting in a single SdH frequency [cf. Fig.~\ref{fig:R-point-explanation}\,(b)]. Upon field rotation the bands continue to cross at the NPs with orthogonal wave functions and thus without interaction, giving rise to a \emph{single} non-dispersive SdH branch, in contrast with experiment.

Shown in Figure~\ref{fig:R-point-explanation}\,(c) is the behavior when including SOC but neglecting the NPs as proposed in Ref.\,\onlinecite{Wang-dHvA-PRB-2020}. Again, a small branch splitting is expected close to [100]. Due to SOC, four distinct FS sheets would exist around R with pairwise degenerate extremal cross-sections for $B\parallel [100]$ that split into four branches under field rotation. Since the extremal orbits on the intermediate sheets 2 and 3 would come close to each other at up to $12$ breakdown junctions (yellow circles) up to 8192 (highly degenerate) breakdown branches are expected as indicated by the shading between branches 2 and 3. None of this is observed experimentally.

The presence of both SOC and NPs are, finally, shown in Fig.~\ref{fig:R-point-explanation}\,(d). Since the spin degeneracy is lifted and the extremal orbits reside on a NP (cyan) for $B\parallel [100]$, two extremal cross-sections and thus two SdH frequencies are expected. Under field rotation these split into four orbits that intersect at the NPs and remain pairwise degenerate. Magnetic breakdown occurs at the same junctions as in Fig.\,\ref{fig:R-point-explanation}\,(c). However, due to the band connectivities only up to $6$ breakdown junctions are expected for each pair of orbits [Fig.~\ref{fig:R-point-explanation}\,(d)]. Calculating the frequency spectra in DFT \cite{supp} we find two dominant branches with a difference between $80$\,T and $90$\,T and a very low dispersion of $\sim10$\,T, as well as a tiny probability for other breakdown branches, all in excellent agreement with experiment. 

Taken together, we reported a general analysis of the band topology of SG\,198 focusing on CoSi \cite{MnSi_nodal_plane}. The SdH frequencies associated with FS sheets at the R point provide unambiguous evidence a sixfold degeneracy at R even for weak SOC, implying a fourfold degeneracy at $\Gamma$.
\cite{CoSi_Nature_hong_19, RhSi_Nature_hasan_19, 2019-Takane-PRL, 2019-Yuan-Science-Advances, Wu_SdH-CPL-2019, Xu_2019_PRB, Wang-dHvA-PRB-2020}. These degeneracies are part of a topological network comprising, in addition, the topological NPs, the fourfold crossings at M, the enforced Weyl points on screw rotation axes, as well as accidental Weyl points at generic positions in the BZ, in combination satisfying the FDT~\cite{supp,nielsen_no_go}. 
Although the NPs do not give rise to additional Fermi arcs at the surface, they will contribute substantially to the bulk topological responses, such as spin anomalous Hall currents and nonlinear optical responses. Moreover, in the light of our findings it seems difficult to justify the use of strongly simplified Hamiltonians of TQPs in analogy to particle physics.

Since our analysis is based on symmetry it is also applicable to other nonmagnetic representatives of SG\,198, such as PdGa, and can be generalized straightforwardly to other chiral or non-centrosymmetric SGs.
This highlights the need for a complete assessment of the topological charges beyond isolated degeneracies in any material.

\begin{acknowledgments}
We wish to thank J. Knolle and A. Rosch for discussions. M.A.W., A.B. and C.P. were supported by the DFG in the framework of TRR80 (project E1 and E3, project-id 107745057), SPP 2137 (Skyrmionics) under grant number PF393/19 (project-id 403191981), GACR Projekt WI 3320/3-1, ERC Advanced Grants No. 291079 (TOPFIT) and 788031 (ExQuiSid), and Germany's excellence strategy EXC-2111 390814868. A.P.S. thanks the Max Planck-UBC-UTokyo Center for Quantum Materials for support.
\end{acknowledgments}

\end{document}


\setcounter{secnumdepth}{3}
\setcounter{equation}{0}
\setcounter{figure}{0}
\renewcommand{\theequation}{S-\arabic{equation}}
\renewcommand{\thefigure}{S\arabic{figure}}
\renewcommand{\thetable}{S\Roman{table}}
\newcommand\Scite[1]{[S\citealp{#1}]}




 \clearpage
\newpage

\title{Supplementary Material for\\
	"Network of topological nodal planes, multifold degeneracies, and Weyl points in CoSi"}

\author{Nico Huber}
\thanks{These authors contributed equally.}
\affiliation{Physik Department, Technische Universit\"at M\"unchen, D-85748 Garching, Germany}
\author{Kirill Alpin}
\thanks{These authors contributed equally.}
\affiliation{Max-Planck-Institute for Solid State Research, Heisenbergstrasse 1, D-70569 Stuttgart, Germany}
\author{Grace Causer}%
\affiliation{Physik Department, Technische Universit\"at M\"unchen, D-85748 Garching, Germany}
\author{Lukas Worch}%
\affiliation{Physik Department, Technische Universit\"at M\"unchen, D-85748 Garching, Germany}
\author{Andreas Bauer}%
\affiliation{Physik Department, Technische Universit\"at M\"unchen, D-85748 Garching, Germany}
\author{Georg Benka}%
\affiliation{Physik Department, Technische Universit\"at M\"unchen, D-85748 Garching, Germany}

\author{Moritz~M. Hirschmann}
\affiliation{Max-Planck-Institute for Solid State Research, Heisenbergstrasse 1, D-70569 Stuttgart, Germany}
\author{Andreas P. Schnyder}
\affiliation{Max-Planck-Institute for Solid State Research, Heisenbergstrasse 1, D-70569 Stuttgart, Germany}
\email{a.schnyder@fkf.mpg.de}

\author{Christian Pfleiderer}%
\affiliation{Physik Department, Technische Universit\"at M\"unchen, D-85748 Garching, Germany}
\affiliation{MCQST, Technische Universit\"at M\"unchen, D-85748 Garching, Germany}
\affiliation{Centre for Quantum Engineering (ZQE), Technische Universit\"at M\"unchen, D-85748 Garching, Germany}
\author{Marc A. Wilde}
\email{mwilde@ph.tum.de}
\affiliation{Physik Department, Technische Universit\"at M\"unchen, D-85748 Garching, Germany}


\date{\today}

\begin{abstract}
In this supplemental material (SM), we present a description of the DFT methods, a symmetry analysis of SG 198, the band structure topology of CoSi with and without SOC, generic tight-binding models for SG 198, as well as details of the experimental methods, the data analysis, and a specific comparison of our Shubnikov-de Haas spectra with the literature.
\end{abstract}

\maketitle

\section{Band topology analysis}
\label{sec:1}

In this first part of the supplemental material (SM), we report the details of the comprehensive analysis of the band topology of space groug (SG)\,198 and CoSi presented in the main text, considering the band topology with and without spin-orbit coupling (SOC). In particular, we analyse the \emph{global} band topology, i.e., the Chern numbers of all crossings of a given band or, where necessary, the set of bands, verifying that these Chern numbers add up to zero in agreement with the fermion doubling theorem (FDT)~\cite{nielsen_no_go}. As one of the main results of our analysis we find that CoSi, taking SOC into account, exhibits trios of symmetry-enforced topological nodal planes (NPs) of the cubic crystal structure for any set of two successive bands. To the best of our knowledge, the implications of these topological NPs have not been considered in the literature.
 
Without SOC we find that CoSi exhibits two trios of NPs close to $E_{\rm F}$. These trios of NPs are not required to be topological by symmetry. However, in our density-functional theory (DFT) calculations we find that the lower trio of NPs, formed by bands~1 and~2, supports a finite topological charge with a Chern number of $\tilde{\nu}^1_{\mathrm{npt}} = -4$. In contrast, the upper trio of nodal planes, formed by bands~3 and 4, which also cross the Fermi level, is topologically trivial, i.e., $\tilde{\nu}^3_{\mathrm{npt}} = 0$. It may be helpful to note that the charge of the lower trio of NPs due to accidental band crossings in the bulk could, in principle, vanish as well. However, our DFT calculations show that this is not the case.

Including SOC, the two trios of NPs exhibit a spin splitting close to $E_{\rm F}$ and decompose into four trios of NPs which are twofold degenerate. We find that these trios of spin-non-degenerate NPs are topological by symmetry alone, because the Weyl points at $\Gamma$ carry an odd topological charge for all bands. This topological charge of the Weyl points at $\Gamma$ can only be compensated by the trios of NPs, since all other accidental or enforced Weyl points contribute an even charge (see Ref.\,\onlinecite{MnSi_nodal_plane} for pedagogical details). Indeed, in our DFT calculations we find that these four trios of NPs carry the charges $(\nu^1_{\mathrm{npt}},\nu^3_{\mathrm{npt}},\nu^5_{\mathrm{npt}},\nu^7_{\mathrm{npt}}) = ( -1,-13,-5,-1)$.

The presentation of the details of our analysis of the band topology is organized as follows. In Sec.~\ref{sec_convention} we specify the conventions for labeling the bands and the Chern numbers of the Weyl points and NPs. In Sec.~\ref{calc_top_char} we present details of the DFT calculations, in particular, how the Chern numbers are directly obtained from the DFT wavefunctions. In Sec.~\ref{sec_sym_analysis_SG_198} we discuss the symmetries of SG 198 and show how they lead to symmetry-enforced band crossings. In Secs.~\ref{sec_CoSI_no_SOC} and~\ref{sec_CoSI_with_SOC} we analyze the global band topology of CoSi without and with SOC, respectively.  Finally, in Sec.~\ref{sec_tight_binding_model} we present two generic tight-binding models, which illustrate the generic symmetry-enforced band crossings in materials with SG198. 
 

\subsection{Conventions and DFT methods}
\label{sec_convention}

\paragraph{Naming conventions  ---}
We denote the Chern number of a nodal point or a nodal plane in a band structure with and without SOC by $\nu^n_A$ and $\tilde{\nu}^n_A$, respectively. Here, the subscript $A$ labels the location of the nodal point in the Brillouin zone or the trio of nodal planes (${\rm npt}$). The superscript $n$ designates the index of the lower band (in energy) that participates in the degeneracy. For example, the Chern number of a two-fold Weyl point at $\Gamma$ formed by the bands $m$ and $m+1$ is denoted by $\nu^m_{\Gamma}$. Further, in the following figures, the Chern numbers of a crossing are indicated by a list of numbers in \{...\} brackets, where the listed Chern numbers correspond from left to right to the bands participating in the given crossing from low to high energy. In this notation, all bands with undefined Chern numbers are skipped.

The relevant bands close to the Fermi energy are enumerated going from low to high energies starting at roughly $-0.5$~eV. For those cases where the band index is not sufficient to identify a crossing unambiguously, we add an additional index enumerating the crossings from low to high energy and from small to large $k$.  For example, with SOC there are two crossings on each $\Gamma$--X line between the bands 2 and 3 as well as 3 and 4. We denote their chiralities by $\nu^2_{\Gamma\text{X},1}$ and $\nu^2_{\Gamma\text{X},2}$ as well as $\nu^3_{\Gamma\text{X},1}$ and $\nu^3_{\Gamma\text{X},2}$. Finally, for simplicity we do not include the spin degeneracy when labeling bands without account of SOC. However, we consider the spin degeneracy when stating the chiralities (Chern numbers) of the crossings.

\paragraph{DFT Methods  ---}
To compute the bands and Chern numbers of CoSi we used WIEN2k~\cite{WIEN2k} and Quantum Espresso with PAW pseudopotentials~\cite{DALCORSO2014337} and the PBE functional~\cite{PBE_functional}. For our purpose it proved to be most efficient to compute the Berry curvature and the Chern numbers using directly the DFT wave functions. Further details are described in Secs.~\ref{calc_ChernNos} and~\ref{calc_charge_nodal_plane}. In Sec.~\ref{alternative_methods} we briefly mention two alternative approaches to compute the Chern numbers of the nodal planes, which are based on tight-binding models derived from maximally localized Wannier functions (MLWF) and the winding of Wannier charge centers, respectively. In this section we  explain also why these approaches are less efficient for our purpose than those described in Sec.~\ref{calc_charge_nodal_plane}.

\subsection{Calculation of topological characteristics} 

\label{calc_top_char}
\subsubsection{Calculation of Chern numbers in DFT} 
\label{calc_ChernNos}

Chern numbers of crossing points were calculated directly using the DFT code Quantum Espresso\,\cite{giannozzi2009quantum}. To do so, crossings on high-symmetry lines were located by dense sampling, where a threshold proportional to the mean change of energy around the minima of the bandgap between two energy levels was used to distinguish between crossings and anti-crossings. To calculate the Chern number of the crossing for a single band $n$, we integrated the Berry curvature by means of a watertight mesh of plaquettes around this crossing, i.e., typically a sphere with a radius smaller than the minimal k-space separation between two crossings, here $0.004\pi$. A Wilson loop around one of the plaquettes is given by\,\cite{fukui_JPSJ_05}
\begin{equation}
\phi_{p,n} = -\text{Im}\ln\left(\prod_{i=1}^4\braket{u_{k_{i,p},n}|u_{k_{i+1,p},n}}\right) , 
\label{wilson}
\end{equation}
with the corners $k_{i,p}$ of the plaquette $p$ and $k_{5,p}=k_{1,p}$ and $u_{k,n}$ given by Bloch's theorem $\Psi(r)_n=\exp(\upi kr)u(r)_{k,n}$. 

The branch of the complex logarithm was chosen to be the negative real axis, so that $|\phi| \ll \pi$ if the plaquettes were sufficiently small and well separated from divergences of the Berry curvature. Further, if the separation of $k_{i,p}$ approaches zero, then $\phi_{p,n}$ approaches the component of the Berry curvature orthogonal to the plaquette, $p$. The Chern number of a single band $n$ is then given by
\begin{equation}
\nu^n_A = \frac{1}{2\pi} \sum_p \phi_{p,n}.
\label{chern}
\end{equation}{
Using the Wannier90\,\cite{pizzi2020wannier90} interface of Quantum Espresso, arbitrary overlap-integrals were calculated of the form
\begin{equation}
M_{mn}(k_1,k_2)=\braket{u_{k_1,n}|u_{k_2,m}}
\end{equation}
between two wave functions of bands $n$ and $m$ and at $k$-points $k_1$ and $k_2$. By evaluating $M_{nn}$ for all plaquettes, Eq.~\eqref{chern} was used to calculate $\nu^n_A$ directly from a DFT calculation.

For the evaluation of the Chern number of double Weyl points at the nodal plane, Eq.~\eqref{wilson} could not be used, since the integration area crosses the nodal plane and a zero bandgap to a band other than $n$ is inevitable. To avoid this, a non-Abelian Wilson loop was used \cite{fukui_JPSJ_05}
\begin{equation}
\Phi_{p,n} = -\text{Im}\ln\det\left(\prod_{i=1}^4 m(k_i,k_{i+1})\right), 
\label{wilson_na}
\end{equation}
where $m(k_i,k_{i+1})$ are the diagonal submatrices of $M(k_i,k_{i+1})$. The total topological charge of the bands included in the submatrices $m(k_i,k_{i+1})$ could then be calculated the same way as before, i.e., using Eq.~\eqref{chern}, choosing $2 \times 2 $ matrices for $m(k_i,k_{i+1})$ comprising the two bands forming one of the nodal planes. Around each crossing it was possible to find a mesh of plaquettes such that there was a nonzero bandgap between the nodal planes. In this way the resulting Chern number represents a topological charge between the two pairs of bands.

\subsubsection{Calculation of the charge of nodal planes in DFT} 
\label{calc_charge_nodal_plane}

To compute the Chern number of the nodal planes in CoSi directly, an integral of the Berry curvature of the area of a cube slightly smaller than the BZ boundary was carried out. As described in the previous section, this integral of the Berry curvature is in practice a plaquette summation. Performing this summation, the separation between the integration cube and the BZ boundary must be smaller than the distance between the closest crossing to the BZ boundary and the boundary itself. However, if the integral of the Berry curvature is carried out the same way as in the previous section, i.e., by a uniform mesh across the cube, one finds that even for high mesh densities $|\phi| \ll \pi$ is no longer given. Increasing the maximal possible mesh size taking into account symmetries of the Berry curvature, i.e., the three effective mirror symmetries as given by the combination of time-reversal with a twofold rotation and the threefold rotation, the reduction of integration area to one quarter of the cube face is still insufficient to guarantee $|\phi| \ll \pi$ since crossings close to the cube, for example double Weyl points, represent divergences of the Berry curvature close to the integration area. 

We addressed this issue by subdividing the plaquettes in which $|\phi| \ll \pi$ is violated into four smaller plaquettes for a threshold of $|\phi| < \pi / 10$ using such an adaptive $k$-point mesh similar to a quad-tree by continuously evaluating new $k$-points and their overlap with other existing points. This method of integration was first described in Ref.\,\onlinecite{gosalbez2015chiral}, where it was applied to a tight-binding model instead of the evaluation using DFT applied in our analysis. Special care was taken where two plaquettes of different sizes met, since the gap between them was not necessarily watertight regarding the Berry curvature. Such gaps were filled by a very slim triangle, with a long side touching the large plaquette and two short sides touching the small plaquettes. Keeping with this procedure no flux was lost when subdividing plaquettes. In particular, we confirmed that after seven levels of subdivision, none of the plaquettes violated $|\phi| < \pi / 10$ for the first 12 bands in the vicinity of the Fermi energy of CoSi.

\subsubsection{Identification of generic Weyl points}
\label{sec_gen_Weyl_points}

Using the above method in our analysis of CoSi we found a mismatch of the Chern numbers of the nodal planes as directly calculated via DFT and the values inferred from the crossings at high-symmetry points / lines via the fermion doubling theorem. We could attribute these differences in the Chern numbers to the existence of generic Weyl points, i.e., Weyl points away from any high-symmetry points or lines. The binary search method we used to identify these generic Weyl points may be summarized as follows.

We started by setting a stepsize variable $s$ to $L/4$, where $L$ was the length of the BZ. We then determined the Chern number of a cube in the BZ with a side length of $L/2$ centered at $\Gamma$ using the method described in Sec.\,\ref{calc_charge_nodal_plane} above. The resulting Chern number was compared to the sum of all Chern numbers of crossings inside the cube. If there was a mismatch, we reduced the side length of the cube by $s$. If the values of the Chern numbers matched, the side length was increased by $s$. Next the stepsize $s$ was halved and the process started again for the new cube, repeating the procedure until $s$ was smaller than some threshold.

In this way, the distance of a generic Weyl point to the resulting cube should be within $s$. Since our method of integrating the cube used an adaptive mesh and therefore samples with a high density at divergences of the Berry curvature on the cube, we could use the sampled mesh to find the exact position where the generic Weyl point is closest to the cube, since the proximity to the Weyl point results in large absolute values of the Berry curvature. Technically this was done by a simple minimum search of the band gap on the adaptive mesh. The resulting $k$-point, $K$, in the BZ will be in close proximity to the generic Weyl point, controlled by the threshold of the binary search and the sampling density of the adaptive mesh. To confirm the existence of the newly found generic Weyl point, we integrated a sphere centered at $K$ with a radius set to the error of the generic Weyl point search. If a nonzero Chern number was calculated, the presence of the generic Weyl point was confirmed and it was checked, if the newly found Weyl point resolved the Chern number mismatch with the nodal planes. If the Chern number of the generic Weyl point did not resolve the mismatch, the entire process of searching for another generic Weyl point was repeated.

As the only caveat, it is important to note that this search method could miss pairs of of generic Weyl points with small separation and opposite Chern numbers, since these may only be detected when the cube separates these two generic Weyl points during the search. However, the method is guaranteed to find at least one generic Weyl point, if there is a mismatch with the Chern number of the nodal plane to start with, since at some point between a cube with zero volume and a cube spanning almost the whole BZ, the Chern numbers of the cube must switch between a mismatch, which is guaranteed to happen for the latter cube, and no mismatch, which must be the case for the cube with no volume, as this cube encloses the $\Gamma$ point only such that no generic Weyl points can be present.

\subsubsection{Alternative methods for the charge of nodal planes}
\label{alternative_methods} 

It proves to be instructive to discuss two alternative methods that are commonly used to compute Chern numbers and explain why these methods are less well suited for the calculation of the charge of the nodal planes.

\paragraph{Tight-binding models derived from MLWF ---}
An alternative method to compute the Chern numbers is to start from a DFT-derived tight-binding (TB) model for which Wilson loops can be computed and summed up. In this approach the TB model is extracted from a subset of DFT bands by means of Maximally Localized Wannier Functions (MLWF), e.g., as implemented in Wannier90 \cite{pizzi2020wannier90}. Here the MLWFs form the basis of the TB model on which the DFT wave functions are projected. 

A well-known advantage of DFT-derived TB models concerns that eigenstates and eigenvalues typically may be computed faster than in DFT, depending on the size of the TB Hamiltonian only, i.e., the number of bands needed to represent a given set of orbitals of some material. Moreover, pertinent information about the chosen DFT bands is relatively easily available in the TB models. 

A main disadvantage concerns, in contrast, that the TB models obtained from the MLWFs frequently do not exactly reproduce the DFT bands, as it is difficult to obtain full convergence in the construction of the MLWFs. Moreover, within Wannier90 the MLWFs for systems with SOC do not fully respect the symmetries of the original system. While these inaccuracies may be ignored in many cases (e.g., for the calculation of surface states), they affect the calculation of the Chern numbers of the nodal planes critically, because the imprecisions and symmetry-breaking of the MLWFs result in a small splitting of the nodal planes causing artificial divergences of Berry curvature at the BZ boundary (see Supplementary Information S2 in Ref.~\onlinecite{MnSi_nodal_plane}). 
Due to these artificial divergences, a Berry curvature integration is not possible and does not even make sense, since the nodal plane is no longer present. 

In order to increase the precision of the MLWF-based TB Hamilonians, a larger number of DFT bands could be included. However, with an increasing number of bands the numerical cost for the fit and the evaluation of the resulting TB model increases considerably and exceeds rapidly a direct integration of the Berry curvature via DFT as discussed in Sec.~\ref{calc_charge_nodal_plane}.   

A possible way of circumventing artificial divergences in the Berry curvature is to explicitly impose the symmetries of the system on the MLWFs, as described in Ref. \,\onlinecite{sakuma2013symmetry}. However, we found it difficult to implement this procedure, as it can lead to changes of the band topology of the TB bands as compared to the DFT bands. On a different note, another option may be to use Wannier90 to first derive a fully symmetric TB model without SOC and then add SOC by hand, as based on a fit to the DFT bands. Unfortunately this proved to be impractical, as the number of SOC parameters was too large to obtain a good fit.

\paragraph{Winding  of Wannier charge centers ---}
The Chern numbers may also be calculated from the winding of Wannier charge centers (WCC), as obtained in DFT and implemented in Z2Pack \cite{gresch2017z2pack}. However, for the determination of Chern numbers of nodal planes addressed in our study this method proved to be inefficient, due to the divergences at the nodal planes, requiring a high sampling rate along an entire line of $k$-points for a single WCC point. In other words, using WCCs, $k$-points must be sampled with a high density even far away from the divergence. Due to this high sampling rate we found a very slow convergence of the WCC calculation making it effectively unfeasible. In comparison, the adaptive mesh described in Sec.~\ref{calc_charge_nodal_plane} proved to be much more efficient, significantly reducing the computation time, such that the calculation of the Chern numbers of the nodal planes became viable.

\subsection{Symmetry analysis for space group 198}
\label{sec_sym_analysis_SG_198}

Before analyzing the details of the CoSi band structure without and with SOC in the next two sections, we discuss in the following the symmetry-enforced band crossings and accidental band crossings for all non-magnetic representatives of the cubic SG\,198 (P$2_13$). Besides CoSi, this includes RhSi, RhGe, CoGe, PdGa, and others. That is, we investigate how the symmetries of SG\,198 together with time-reversal symmetry enforce and constrain possible band crossings at and away from high-symmetry points and lines. In particular, (i) we infer the possibility of band crossings at high-symmetry points/lines from the dimensionality of the irrpes, (ii) we discuss symmetry-enforced band crossings on lines that are left invariant by screw rotations, and (iii) we study the multiplicity of accidental band crossings. 

\paragraph{(i) Band crossings on high-symmetry points and lines ---}%
The existence of band crossings on high-symmetry points and lines follows from the dimensionality of the irreps of the little groups. We obtained the dimensionality of these irreps from the Bandrep tool on the Bilbao Crystallographic Server~\cite{bradlyn_Nature_bandrepTool}. The result is shown in Table~\ref{TabdimIrreps}, which lists the dimensionality of all irreps of all little groups on points and lines,  both for the spinless case (no SOC) and the spinful case (with SOC). For example, without SOC we find that there are irreps of dimension one, two, and three at $\Gamma$. Thus, there can be two-fold and three-fold degeneracies at  $\Gamma$. 

Similar arguments apply for the other entries. The entries written in italics correspond to point crossings. It is interesting to note that there is no point crossing at X, since X is part of a nodal plane. Also, all two-fold degeneracies along lines are part of nodal planes, such that there is no line crossing. Hence, in order to infer the topology of the nodal planes it is sufficient to look at the entries in italics, representing the only entries that could potentially lead to topological charges that can only be compensated by the nodal planes (see Secs.~\ref{sec_CoSI_no_SOC} and~\ref{sec_CoSI_with_SOC}).
 
\begin{table}[t!]
\centering
\begin{tabular}{|c|c|c|}
\hline 
k point & no SOC & with SOC \\ 
\hline 
$\Gamma (0,0,0)$ & 1d or \emph{2d}, \emph{3d} & \emph{2d}, \emph{4d} \\ 
\hline 
X$(0,\pi,0)$ & 2d & 2d(2) \\ 
\hline 
M$(\pi,\pi,0)$ & 2d(2) & \emph{4d} \\ 
\hline 
R$(\pi,\pi,\pi)$  & \emph{4d(2)} & 2d, \emph{6d} \\ 
\hline 
\hline 
k path & no SOC & with SOC \\ 
\hline 
$\Gamma$-X & 1d(2) & 1d(2) \\ 
\hline 
$\Gamma$-M & 1d & 1d \\ 
\hline 
$\Gamma$-R & 1d(3) & 1d(3) \\ 
\hline 
M-R & 2d(2) & 2d(2) \\ 
\hline 
X-M, X-R & 2d & 2d \\ 
\hline 
\end{tabular} 
\caption{
\label{TabdimIrreps} 
Dimensionality of the irreducible representations (irreps) of the little groups of SG\,198 with time-reversal symmetry as listed at the Bandrep tool on the Bilbao Crystallographic Server~\cite{bradlyn_Nature_bandrepTool}. The number in brackets represents the number of distinct irreps. The entries written in italics correspond to point crossings. We note that the case without SOC corresponds to spinless electrons, while for the case with SOC we consider spinful electrons, i.e., double groups. }
\end{table}

\paragraph{(ii) Symmetry-enforced movable crossings ---}
From the compatibility relations between irreps the existence of movable crossings that are symmetry-enforced by non-symmorphic symmetries may be inferred. Since SG\,198 contains two-fold screw rotations but no (glide) mirror or inversion, these movable crossings can only exist as point nodes on screw rotation axes. We find that without SOC no movable crossings are enforced, while with SOC there are movable crossings on the $\Gamma$-X path, which are enforced due to the compatibility relations between the 2d irreps at $\Gamma$ and X. However, these movable crossings are at different band indices than the nodal planes. Hence, unlike for SG\,19.27~\cite{MnSi_nodal_plane}, the movable crossings on $\Gamma$--X  in SG\,198 do not enforce nontrival topological charges of the nodal planes. Instead, in SG\,198 the pinned crossing at $\Gamma$ ensures the non-vanishing charge of the planes, see Sec.~\ref{sec_CoSI_with_SOC} below.

\paragraph{(iii) Multiplicity of accidental band crossings ---}
To infer the global band topology in the entire Brillouin zone (BZ), accidental band crossings at and away from high-symmetry lines or planes must also be considered. By virtue of the fermion doubling theorem, these accidental crossings are important for the precise topological charge of the nodal planes. Therefore, we determined the multiplicity of these accidental crossings, computing for a given accidental crossing the number of symmetry-related copies with the same topological charge (Chern number).  It transpires that the multiplicity of these accidental crossing points is independent of the presence or absence of SOC. Moreover, symmetry-related Weyl points in SG\,198 always carry the same chirality, i.e., the same sign of the Chern number, since SG\,198 does not contain inversion or mirror symmetries. 

We find that accidental Weyl points on the twofold axes  $\langle 100 \rangle$ have multiplicity 6, i.e., they come in 6 copies, which are related to each other by time-reversal symmetry and threefold rotation. The same is true for accidental crossings on the twofold rotation axes within the nodal planes (i.e., at the edges of the BZ cube), whose degeneracies are larger than twofold. On the threefold axes (body diagonals of the cube) the accidental crossings have multiplicity 8, i.e., there are symmetry-related crossings in each of the 8 BZ octants. Furthermore, we note that accidental crossings are located on the effective mirror planes $k_{x,y,z}=0$, which are left invariant by the effective mirror symmetry formed by the combination of a twofold screw rotation with time-reversal symmetry (cf.~\cite{MnSi_nodal_plane}). A Weyl point on such an effective mirror plane has multiplicity 12, i.e., it appears in 12 copies due to the other effective mirror symmetries and the threefold rotation. Finally, Weyl points at generic positions $(k_x,k_y,k_z)$, away from any high-symmetry lines or planes, have multiplicity 24, i.e., they have symmetry-related copies at $(\pm k_x, \pm k_y, \pm k_z)$ (due to rotation and time-reversal symmetry) and at the coordinates where the $k_i$ are permuted by the threefold rotations.

From the above analysis of accidental crossings, it follows that a possible symmetry-enforced topological charge of the nodal planes must be \emph{odd}. If it were even, one could put accidental crossings on the twofold and threefold axes with opposite topological charges, yielding a net extra charge of $2 \mathbb{Z}$, such that the charge of the nodal plane is canceled.  

\subsection{C\MakeLowercase{o}S\MakeLowercase{i} without spin-orbit coupling}
\label{sec_CoSI_no_SOC}

Following the symmetry analysis of the band topology of general band structures in SG\,198, we now address specifically the properties of CoSi. The strength of SOC in CoSi is on the order of $\sim 10$~meV\added{,} and thus relatively small and rather difficult to resolve by many experimental techniques, notably angle-resolved photoemission spectroscopy (ARPES). In turn, it is instructive to analyse at first the band structure of CoSi without SOC and study next how the analysis changes in the presence of SOC.  

In this spirit it is essential to note that without SOC it is possible, in principle, to have a non-zero topological charge of the nodal planes. For example, there can be a crossing point at $\Gamma$ with charge $\tilde{\nu}_\Gamma = +4$ that is compensated by a nonzero charge of the nodal plane. However, $\tilde{\nu}_\Gamma = +4$ can also be compensated by Weyl points on $\Gamma$--X and $\Gamma$--R with charges $+2$ with multiplicity 6 and $-2$ with multiplicity 8, respectively. Taken together, without SOC it is possible to have topological nodal planes in principle. However, their existence is not guaranteed by symmetry alone.

\begin{figure}[t!]
\centering
\includegraphics[width = 1 \columnwidth]{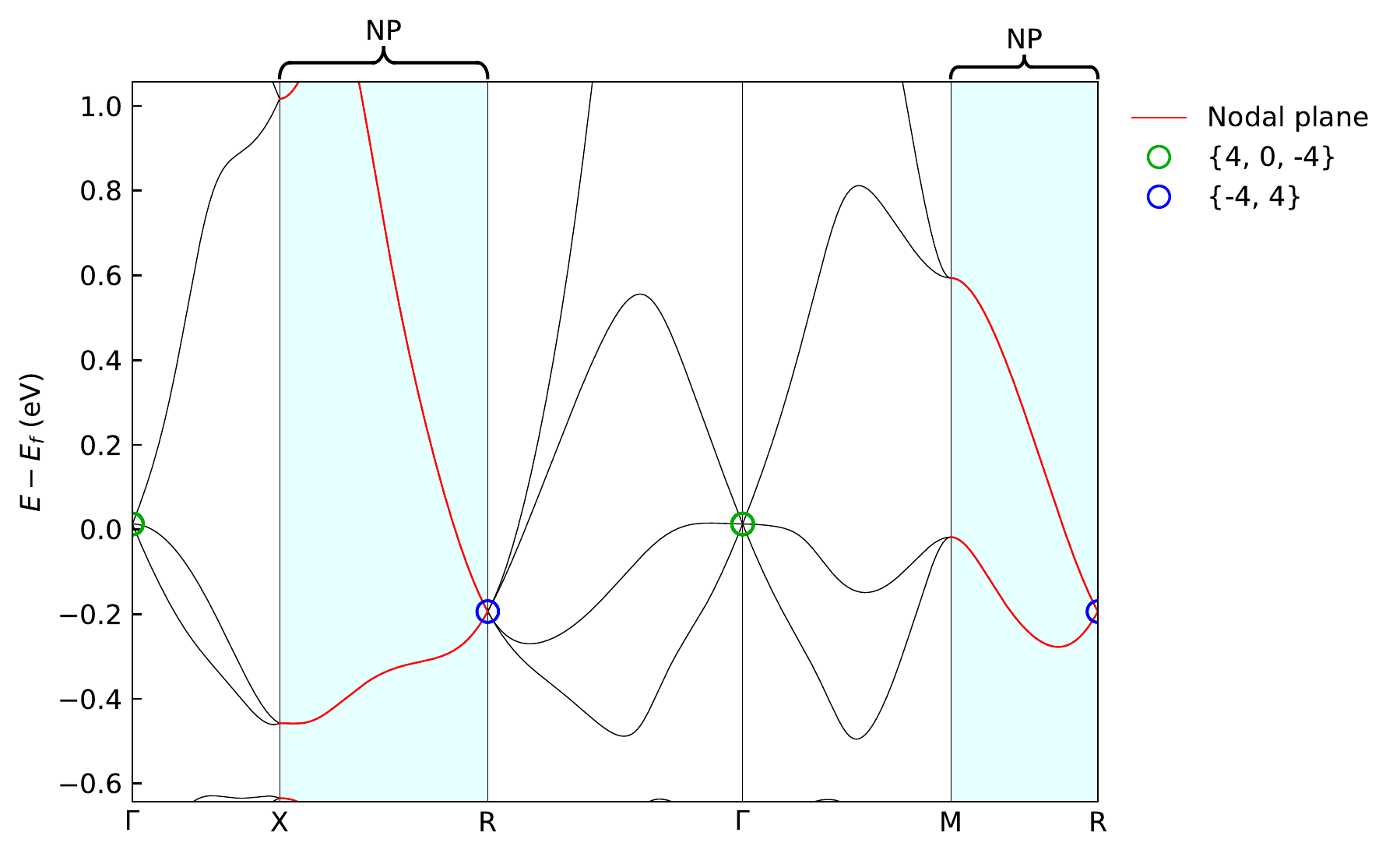}
\caption{\label{fig_DFT_noSOC}
DFT band structure of CoSi without SOC. Band degeneracies are marked by different colors. The Chern numbers of these crossing are denoted by curly brackets $\{..\}$. See text for details of notation. The trios of topological NPs are marked in red.
}
\end{figure}

Shown in Fig.~\ref{fig_DFT_noSOC} is the band structure of CoSi without SOC near $E_{\rm F}$ along high-symmetry lines. We observe that there are point crossings at $\Gamma$ (threefold) and R (fourfold). In the energy range shown neither accidental crossings  (cf.~Table~\ref{TabdimIrreps}) nor movable symmetry-enforced crossings exist, consistent with the analysis presented in Sec.~\ref{sec_sym_analysis_SG_198}. The threefold degeneracy at $\Gamma$, supports the charges $(\tilde{\nu}^1_\Gamma,\tilde{\nu}^2_\Gamma,\tilde{\nu}^3_\Gamma) = (+4,0,-4)$, while the fourfold point at R yields $(\tilde{\nu}^2_{\mathrm{R}},\tilde{\nu}^4_{\mathrm{R}}) = (-4,+4)$. To simplify the discussion presented in the following, we assume that there are no further accidental crossings among the bands of Fig.~\ref{fig_DFT_noSOC}, consistent with our DFT calculations. For this assumption we found that the topological charges of the three bands compensate each other as follows:

\begin{itemize}

\item
The bottom band, $n=1$, carries a charge of +4 at $\Gamma$, which cannot be compensated by the fourfold point at R, because its chirality is ill defined for $n=1$. This is due to its degeneracy with $n=2$ and because it is part of the nodal plane. That is, the Chern number cannot be defined for $n=1$ around R, because R cannot be enclosed by a gapped manifold due to the nodal planes.
Hence, the charge at $\Gamma$ must be compensated by the only other nodal feature present, namely, the nodal plane, which must have charge $\tilde{\nu}^1_{\mathrm{npt}} = -4$.

\item
The second band, $n=2$, carries a trivial charge at $\Gamma$. Thus, the lowest two bands in total support a charge $\tilde{\nu}^1_\Gamma + \tilde{\nu}^2_\Gamma = 4 + 0 = 4$. The nodal plane, on the other hand, has a compensated charge for the lowest two bands, because it is completely filled. For the lowest two bands, $n=1$ and $n=2$, the fourfold point at R features a well defined charge of $\tilde{\nu}^2_{\mathrm{R}} =  -4$, i.e., the opposite of the charge at~$\Gamma$. As a consequence, the charges at $\Gamma$ and R compensate each other, such that there are two spin degenerate Fermi arcs connecting $\Gamma$ to R in the BZ of the surface.

\item
For a filling just above the third band, $n=3$, we find that the charges at $\Gamma$ cancel, while the fourfold point at R has no well defined charge. Thus, the upper nodal plane, formed between bands $n=3$ and $n=4$, does not need to cancel anything, and is trivial in the absence of accidental crossings, i.e.,  $\tilde{\nu}^3_{\mathrm{npt}} = 0$.
\end{itemize}

In conclusion, without SOC the lower nodal plane is charged with $\tilde{\nu}^1_{\mathrm{npt}} = -4$, while the upper nodal plane is trivial, i.e., $\tilde{\nu}^3_{\mathrm{npt}} = 0$. This conjecture is in perfect agreement with the direct calculation of the Chern numbers of the nodal planes, using the method described in Sec.~\ref{calc_charge_nodal_plane}. Moreover, from our DFT calculations we find that the nodal planes at higher energies have the charges
$(\tilde{\nu}^5_{\mathrm{npt}}, \tilde{\nu}^7_{\mathrm{npt}}, \tilde{\nu}^9_{\mathrm{npt}},\tilde{\nu}^{11}_{\mathrm{npt}}) = ( 28,-24, 8, 80)$. It is remarkable that some of the trios of nodal planes in CoSi carry such high Chern numbers, which are likely to be among the largest reported ever for any real material. 

\subsection{C\MakeLowercase{o}S\MakeLowercase{i} with spin-orbit coupling}
\label{sec_CoSI_with_SOC}

We turn now to the properties of CoSi when taking into account the effects of SOC. This corresponds to the properties of the real material as investigated experimentally by means of Shubnikov-de Haas spectroscopy reported in the main text. We note that the SdH data are sensitive to SOC-induced band splitting. The SOC renders the band topology of CoSi considerably more complicated, as it increases the number of bands and leads to numerous accidental crossings between the spin-split bands. Therefore, on the one hand, the results discussed in the previous section are of limited use, as the additional accidental crossings change the topological charges of the nodal planes. On the other hand, since previous experimental studies reported in the literature were analyzed without account of the effects of SOC and the presence of nodal planes, the analysis summarized in Sec.\,\ref{sec_CoSI_no_SOC} above represents an important point of reference.

In stark contrast to the band structure without SOC and as the main difference, the presence of SOC enforces a non-zero topological charge of the nodal planes by symmetry alone~\cite{MnSi_nodal_plane,2019_PRB_Yuxin}. This differences originates from a fourfold point or a single Weyl point at $\Gamma$ with an odd topological charge (see Table~\ref{TabdimIrreps}), which can only be compensated by the trio of nodal planes. Namely, all other possible band crossings contributing to the charge of the NPs have even multiplicity, which cannot cancel the odd contribution from the crossing at $\Gamma$. Hence, with SOC all trios of nodal planes must be topological.

\paragraph{Band crossings \& Chern numbers in Quantum Espresso  ---}

\begin{table}[t!]
\centering
\begin{tabular}{|c|c|c|c|c|}
\hline 
Band index & $\Gamma$ & M & R & NP \\ 
\hline 
1 & +1 & - & -  & -1  \\ 
\hline 
2 & -1 & -2 & - & +1   \\  
\hline 
3 & +3 & - & - & -13   \\ 
\hline 
4 & +1 & +2 & -4  & +13  \\ 
\hline 
5 & -1 & - & -  & -5  \\ 
\hline 
6 & -3 & +2 & 0  & +5  \\ 
\hline 
7 & -3 & - & -  & -1  \\ 
\hline 
8 & -1& -2 & +4  & +1  \\ 
\hline 
\end{tabular} 
\caption{\label{Tab_Chiralities_At_TRIMs}
Topological charges, i.e., Chern numbers, obtained from the DFT calculations for the eight bands of CoSi with SOC above the global band gap\added{,} between roughly $-0.5$~eV and 1~eV. The second, third, and fourth columns give the topological charges of the point crossings at $\Gamma$ (twofold or fourfold), at M (fourfold), and at R (sixfold), respectively. As R and M are situated on twofold degenerate NPs, their Chern numbers are calculated with the non-Abelian Berry curvature for every other band index, since for odd fillings the Chern number is not well defined. The fifth column gives the topological charge of the trio of NPs, as computed from the DFT (see Sec.~\ref{calc_charge_nodal_plane}). 
}
\end{table}
 
We now address all topological crossings on high-symmetry points and along high-symmetry lines for the eight bands between -0.5~eV and 1~eV above the global gap as shown in Figs.~\ref{fig_DFT_withSOC} and~\ref{fig_DFT_withSOC_chiralities}. The bands are numbered consecutively from low to high energy, starting from $n=1$ for the band above the global band gap at $-0.5$~eV. We used DFT to compute the Abelian and non-Abelian Chern numbers of these crossings directly and check for each band, whether the topological charges (Chern numbers) cancel out in the entire BZ, as required by the Nielsen-Ninomiya fermion doubling theorem. If the charges of the crossings on high-symmetry points and lines do not cancel out for a given filling,  there must be additional uncompensated accidental crossings, somewhere away from the high-symmetry lines, such that the doubling theorem is satisfied. 

While pursuing such an assessment, it is important to note that the fermion doubling theorem must be satisfied for the sum of all topological charges (i.e., Abelian or non-Abelian Chern numbers) that can be computed for a given filling. For example, for the fourfold degeneracy at M formed by the bands 1 to 4, the charge must cancel when bands 1 and 2 are filled, and also when all bands 1 to 4 are filled. However, when only bands 1 (or bands 1 to 3) are filled, the Chern number cannot be computed, since band 1 is degenerate with band 2 (band 3 is degenerate with band 4) on the nodal plane. 

The relevant high-symmetry points that must be considered are the time-reversal invariant momenta (TRIMs) listed in Table~\ref{TabdimIrreps}. The degeneracies at these TRIMs are: twofold and fourfold points at $\Gamma$, fourfold points at M, and sixfold points at R. The Chern numbers of these degeneracies for each of the eight bands (fillings) are listed in Table~\ref{Tab_Chiralities_At_TRIMs}. 

In the following, we consider in addition the accidental crossings on the four high-symmetry lines $\Gamma$--X, $\Gamma$--M, $\Gamma$--R, and M--R. These are shown in Fig.~\ref{fig_DFT_withSOC_chiralities} and their Chern numbers are indicated by the color of the open circles. Adding up the Chern numbers of all of these crossings for each of the eight bands (fillings) and comparing the resulting sum with the topological charge of the trio of nodal planes as computed directly from the DFT (6th column of Table~\ref{Tab_Chiralities_At_TRIMs}) we find:

\begin{figure}[t!]
\centering
\includegraphics[width = 1 \columnwidth]{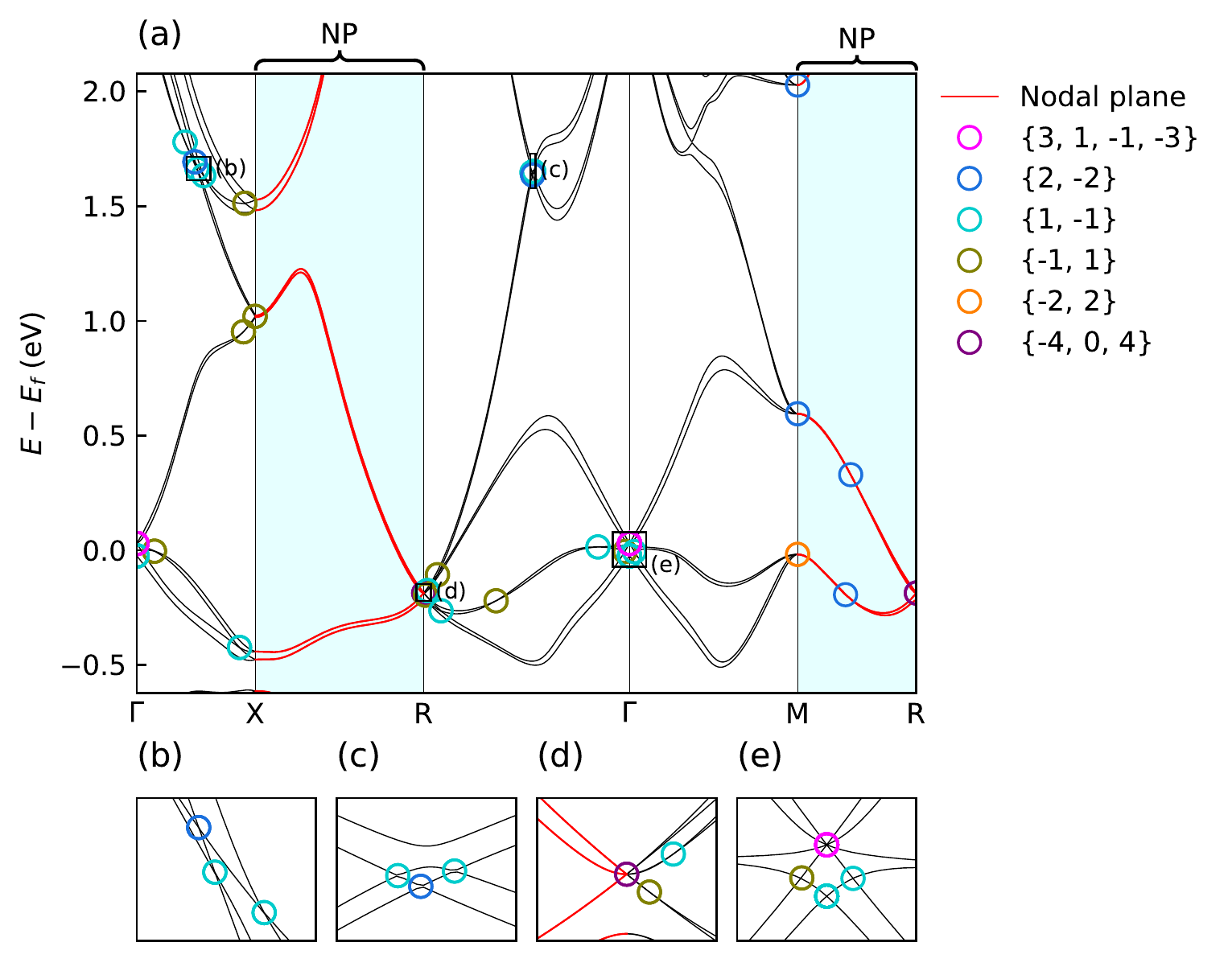}
\caption{\label{fig_DFT_withSOC}
(a) DFT band structure of CoSi with SOC. 
Band degeneracies are marked by different colors. The Chern numbers of these crossing are denoted by curly brackets $\{..\}$ (see text for details of notation). The trio of topological NPs is marked in red.
The insets (b)-(e) show close-up views of crossings (b) along $\Gamma$--X, (c) R--$\Gamma$, (d) at R and (e) at $\Gamma$.  
Further close-up views along the high-symmetry lines are are shown in Fig.~\ref{fig_DFT_withSOC_chiralities}. 
}
\end{figure}

\begin{figure*}[t!]
\centering
\includegraphics[width = 0.8 \textwidth]{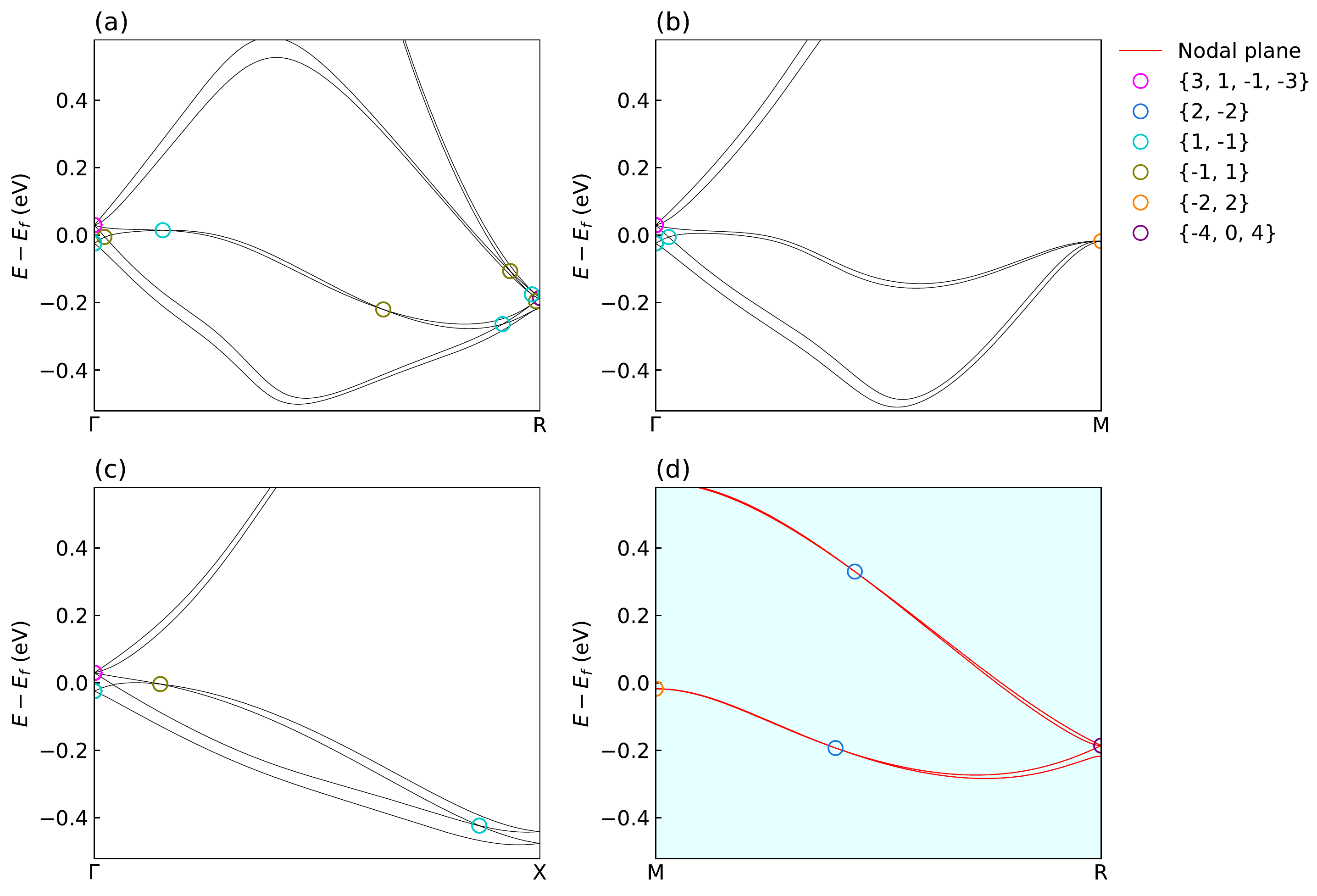}
\caption{\label{fig_DFT_withSOC_chiralities}
(a)-(d) DFT band structure of CoSi with SOC along the high-symmetry lines. (a) $\Gamma$--R, (b) $\Gamma$--M, (c) $\Gamma$--X, and (d) M--R. Band degeneracies are marked by different colors. The Chern numbers of these crossing are denoted by curly brackets $\{..\}$ (see text for details of notation). The trios of topological NPs are marked in red.
}
\end{figure*}

\begin{itemize}
\item 
For band 1 we find that there is only one point crossing, namely the Weyl point at $\Gamma$, which contributes the charge  $\nu^1_\Gamma =  + 1$. Thus due to the fermion doubling theorem the trio of nodal planes must have charge $\nu^1_{\mathrm{npt}} = -1$. This is in perfect agreement with a direct calculation of the charge of the nodal plane (see Table~\ref{Tab_Chiralities_At_TRIMs}), suggesting that there are no other uncompensated accidental crossings between bands 1 and 2. 
For this filling and in general all odd fillings, all possible surface terminations lead, due to the NP trios, to a closed band gap at every point in the BZ, so no Fermi arcs are able to form in between these band pairs, as the bulk-boundary correspondence cannot be applied.
 
\item 
For a filling up to band 2, the nodal plane does not contribute a net topological charge. Therefore, we checked whether the crossings on the high-symmetry points and lines cancel out between them (see Fig.~\ref{fig_DFT_withSOC_chiralities}). The fourfold point at M has multiplicity 3 and charge $-2$. Hence, it contributes $3\nu^2_{\text{M}} = -6$ (see Table~\ref{Tab_Chiralities_At_TRIMs}).  Further, on the  $\Gamma$--X line there is a Weyl point with multiplicity 6 and charge +1, giving $6\nu^2_{\Gamma\text{X},1} = 6$. The Weyl points on the $\Gamma$--R line cancel each other since $ \nu^2_{\Gamma\text{R},1} +  \nu^2_{\Gamma\text{R},2} = -1 + 1 = 0$. On the M--R line there are fourfold crossings with multiplicity 6 and charge +2, yielding $6\nu^2_{\text{MR}} = 12$. Finally, on $\Gamma$--M there are Weyl points with charge +1 and multiplicity 12, giving $12\nu^2_{\Gamma\text{M}} = 12$.  
Adding these contributions, the total charge of all degeneracies on all high-symmetry lines and planes is
$3\nu^2_{\text{M}}  +6\nu^2_{\Gamma\text{X},1}+ 6\nu^2_{\text{MR}} + 12\nu^2_{\Gamma\text{M}} = -6+ 6   + 12  +12 =  24$, which does not cancel. Thus,  there must be a set of additional uncompensated Weyl points away from the high-symmetry points / lines, such that the doubling theorem is satisfied. For example, there could be an additional Weyl point at a generic position in the BZ with charge $-1$ and multiplicity 24, giving $24\nu^2_{\text{generic}} = -24$. Indeed, in our DFT calculations we found an uncompensated Weyl point at $(-0.49,  -0.23 ,  -0.49 ) 2 \pi$ (and symmetry related points) with charge $-1$ and multiplicity 24, which is sufficient to compensate the charges along the high-symmetry lines. Fermi arcs, connecting the here stated charges projected onto the surface BZ, are possible for this filling, if a suitable termination can be found, such that the band gap on paths between these crossings remains open on the surface BZ.

\item
For a filling up to band 3, we find the lowest band of the fourfold point at $\Gamma$ to yield the charge $\nu^3_{\Gamma} = 3$ (see Table~\ref{Tab_Chiralities_At_TRIMs}). On the $\Gamma$--X line there is one Weyl point with charge -1 and multiplicity 6, giving $6\nu^3_{\Gamma\text{X}} = -6$, while on the $\Gamma$--R line there are three Weyl points, two of which cancel each other, yielding  $8\nu^3_{\Gamma\text{R},1} + 8\nu^3_{\Gamma\text{R},2} +8\nu^3_{\Gamma\text{R},3} = -8$ (see Fig.~\ref{fig_DFT_withSOC_chiralities}). Apart from these charges there are no other crossings on high-symmetry lines / planes. Hence, the nodal plane must compensate these charges, i.e., $\nu^3_{\mathrm{npt}} = - \nu^3_{\Gamma} - 6\nu^3_{\Gamma\text{X}} - 8\nu^3_{\Gamma\text{R},1} - 8\nu^3_{\Gamma\text{R},2} - 8\nu^3_{\Gamma\text{R},3}  =  - 3 + 6 + 8  = + 11$. Comparing this value to the value obtained in a direct calculation of the charge of the nodal plane, which is $  \nu^3_{\mathrm{npt}} = -13$ (see Table~\ref{Tab_Chiralities_At_TRIMs}), we infer that there must be an additional accidental Weyl point somewhere at a generic position in the BZ, with charge +1 and multiplicity 24, giving  $24\nu^3_{\text{generic}} = +24$. Indeed, our DFT calculations show that there is an uncompensated Weyl point at $(-0.20, -0.46, -0.47 ) 2 \pi $ with charge of +1 and a multiplicity of 24, which is sufficient to compensate the other charges.

\item
For filling up to band 4, the nodal planes do not contribute any topological charge, as an even number of bands is filled. We have therefore confirmed that the charges of the high-symmetry points and high-symmetry lines cancel each other. For the fourfold point at $\Gamma$ we need to add up the charges from bands 3 and 4, giving $\nu_{\Gamma}^{3} + \nu_{\Gamma}^{4} = 4$, while the sixfold degeneracy at R contributes a charge of $\nu^4_{\text{R}} = -4$. Noting that there are no other crossings on high-symmetry  points / lines that contribute, we find that the sum of these charges vanishes, $\nu_{\Gamma}^{3} + \nu_{\Gamma}^{4} + \nu^4_{\text{R}} = 4-4 = 0$, in agreement with the fermion doubling theorem. The four resulting Fermi arcs connecting $\Gamma$ and R bulk states can be seen in Ref.\,\cite{CoSi_PRL_17}.

\item   
For filling up to band 5, we find that the fourfold point at $\Gamma$ gives a total charge of +3 of bands 3, 4, and 5, i.e., $\nu_{\Gamma}^{3} + \nu_{\Gamma}^{4} + \nu_{\Gamma}^{5} = 3$. On the $\Gamma$--X line there is a movable Weyl point with charge -1 and multiplicity 6, resulting in a charge  $6\nu^5_{\Gamma\text{X}} = -6$. On the $\Gamma$--R line there is a crossing very close to R contributing a charge $8\nu^5_{\Gamma\text{R}} = 8$. Since an odd number of bands is filled, the point crossings at M and R within the nodal plane do not contribute any charge. Hence, the nodal plane must compensate these charges, i.e.,  $  \nu^5_{\mathrm{npt}} = - \nu_{\Gamma}^{3} - \nu_{\Gamma}^{4} - \nu_{\Gamma}^{5} - 6\nu^5_{\Gamma\text{X}} - 8\nu^5_{\Gamma\text{R}}  = - 3 +6 -8 = -5$. Since this value agrees with the value obtained from direct calculations (see Table~\ref{Tab_Chiralities_At_TRIMs}), we conclude that there are no additional uncompensated Weyl points at generic positions within the BZ.

\item 
For filling up to band 6, the sixfold crossing at R contributes $\nu^4_{\text{R}} + \nu^6_{\text{R}} = -4 + 0 = -4$, while the fourfold crossing at M gives  $3\nu^6_{\text{M}} =6$. On the $\Gamma$--X and the $\Gamma$--R lines there are Weyl points with charge -1 and multiplicity 6 and 8, respectively, contributing $6\nu^6_{\Gamma\text{X}} = -6$ and $8\nu^6_{\Gamma\text{R}} = -8$. On the M--R line, on the other hand, one finds two Weyl points with charge -1, giving $6 \nu^6_{\text{MR}}   = 12$. Adding up all of these charges yields $\nu^4_{\text{R}} + \nu^6_{\text{R}}  + 3\nu^6_{\text{M}}  + 6\nu^6_{\Gamma\text{X}} + 8\nu^6_{\Gamma\text{R}} +  6 \nu^6_{\text{MR}} = -4 +6 -6 -8 +12 = 0$, in agreement with the doubling theorem. This suggests the absence of uncompensated Weyl points away from high-symmetry points and lines. Once again, Fermi arcs, connecting different crossings of opposite charge, are possible for this filling.

\item 
For filling up to band 7, there is a second fourfold point at $\Gamma$ with a charge  $\nu_{\Gamma}^{7} = -3$, while the other crossings at TRIMs do not contribute. On the high-symmetry line $\Gamma$--X there are two Weyl points of identical charge +1, contributing $6\nu^7_{\Gamma\text{X},1} + 6\nu^7_{\Gamma\text{X},2} = 12$, while there is a crossing with charge +2 on the  $\Gamma$--R line, giving  $8\nu^7_{\Gamma\text{R}} = 16$. This crossing on  $\Gamma$--R is somewhat unusual, as the charge of a crossing point on a three-fold axis, should either be $\pm 1$ or $\pm 3$~\cite{tsirkin_vanderbilt_PRB_17}. However, it is possible that this crossing consists actually of a Weyl point on the rotation axis with charge -1 and three copies of crossing points with charge +2 very close around the rotation axis. Adding up all of these charges we obtain $\nu_{\Gamma}^{7} + 6\nu^7_{\Gamma\text{X},1} + 6\nu^7_{\Gamma\text{X},2} + 8\nu^7_{\Gamma\text{R}} = -3 + 12 + 16 = 25$, which must be compensated by the charge of the nodal plane, i.e., $\nu^7_{\mathrm{npt}} = -25$. Comparing this to the charge of the nodal plane, as inferred directly from DFT, which is $\nu^7_{\mathrm{npt}} = -1$ (see Table~\ref{Tab_Chiralities_At_TRIMs}), we concluded that there must be additional uncompensated Weyl points away from high symmetry lines. In turn, we performed a Weyl point search using DFT and found four Weyl points at $(-0.14, 0, -0.18)2\pi$, $(-0.06, 0, -0.22)2\pi$, $(0, -0.09, -0.25)2\pi$ and $(-0.05, 0, -0.32)2\pi$ with charges 1, -1, -1 and -1, respectively, and a multiplicity of 12. Thus, the charge of the sum of these additional Weyl points is -24, which reduces the total sum of all charges in the BZ away from the nodal plane to 1. This confirms a charge of the nodal plane of -1, obtained in a direct calculation using DFT.
\end{itemize}

To summarize, when calculating the charges of nodal planes directly using DFT we obtain
$(\nu^1_{\mathrm{npt}},\nu^3_{\mathrm{npt}},\nu^5_{\mathrm{npt}},\nu^7_{\mathrm{npt}}) = ( -1,-13,-5,-1)$. This result is consistent with an independent calculation using the fermion doubling theorem.

We note that these charged nodal planes do not lead to additional Fermi arcs, as there does not exist a full gap on 2D planes for odd fillings, for which the bulk-boundary correspondence could be applied. Nevertheless, the Berry curvature of the nodal planes are expected to give substantial contributions to bulk topological responses, such as, e.g., anomalous spin Hall effects, or nonlinear optical responses. For even fillings, on the other hand, there are Fermi arcs which connect the projected degeneracy points in the surface BZ.

\paragraph{Robustness of band crossings under small perturbations  ---}
\label{sec_CoSI_with_SOC_MinimalPhysicalCrossings}

It is important to emphasize that, ultimately, only those topological band crossings that are robust under small, symmetry-preserving perturbations have a significant effect on the physical responses via their Berry curvature contributions. As several of the Weyl points we identified in our analysis occur in small gaps and very close to the nodal planes, their relevance for the physical properties is not evident. Indeed, we even found that the accidental Weyl points on the $\Gamma$--M line, observed using Quantum Espresso, do not appear in the VASP DFT implementation. This raises doubts about the stability of similar crossings, since many are close to the nodal planes and appear to be rather accidental. However, having obtained specific values of the charges, it is possible to consider manually how these charges are canceled or get gapped out under small, symmetry-preserving perturbations. Moreover, it is possible to track the fate of crossings in the bulk and if they are moved into the nodal planes, as most of them are not symmetry-enforced.

To identify those crossings that are most robust, which in turn are likely to dominate the physical responses, we consider next symmetry-preserving perturbations to remove those crossings that are not enforced by symmetry, and which can be removed by altering a band gap of less than 100~meV. In doing so we will not change the order of the bands at the TRIMs. This will give us a liberal lower bound on the topological charges to be expected among the nodal planes. In this rough assessment we find that the nodal planes remain topological and that the charge of the fourth nodal plane, $\bar{\nu}^4_{\mathrm{npt}}$, is  enhanced by accidental crossings. 

\begin{itemize}
\item 
Band 1 contains no accidental crossings. The chiralities remain $\bar{\nu}^1_\Gamma =  1$ and $\bar{\nu}^1_{\mathrm{npt}} = -1$. 

\item 
Band 2 must contain $6\bar{\nu}^2_{\Gamma\text{X}} = 6$ because adjacent two- and fourfold representations at $\Gamma$ form a variant of an hourglass state on $\Gamma$--X. On $\Gamma$--R the nodal points evidently compensated each other. M--R can be simplified with the fourfold crossings meeting at M such that $3\bar{\nu}^2_{\text{M}} = +6$ remains. The total charge $6\bar{\nu}^2_{\Gamma\text{X}} + 3\bar{\nu}^2_{\text{M}} = 6 +6 = 12 $, indicates that there must still be an accidental crossing with a multiplicity of 12. By changing the sign of one of the chiralities, $\bar{\nu}^2_{\Gamma\text{X}}$ or $\bar{\nu}^2_{\text{M}}$, involved these could be easily removed.

\item
Band 3 does not need the additional crossing to band 4 on $\Gamma$--X. On $\Gamma$--R the three crossings either partially cancel or are close to the nodal plane. The nodal plane must compensate only the point at $\Gamma$ with $\nu^3_{\Gamma} = 3$ thus the nodal plane acquires the charge $\bar{\nu}^3_{\mathrm{npt}} = -\bar{\nu}^3_{\Gamma}  = -3$. 

\item  
Band 4 is unchanged and thus $\bar{\nu}_{\Gamma}^{3} + \bar{\nu}_{\Gamma}^{4} + \bar{\nu}^4_{\text{R}} = 4-4 = 0$ vanishes.

\item
Band 5 at $\Gamma$ gives $\bar{\nu}_{\Gamma}^{3} + \bar{\nu}_{\Gamma}^{4} + \bar{\nu}_{\Gamma}^{5} = 3$. The movable Weyl points on $\Gamma$--X  and on $\Gamma$--R can be moved into the nodal plane. The nodal plane must compensate the charges at $\Gamma$ only, hence $\bar{\nu}^5_{\mathrm{npt}} = -(\bar{\nu}_{\Gamma}^{3} + \bar{\nu}_{\Gamma}^{4} + \bar{\nu}_{\Gamma}^{5}) = -3$. 

\item 
Band 6 allows for no obvious simplifications. The total charge is $\bar{\nu}^4_{\text{R}} +\bar{\nu}^6_{\text{R}} + 6\bar{\nu}^6_{\Gamma\text{X}} + 8\bar{\nu}^6_{\Gamma\text{R}} +  6\bar{\nu}^6_{\text{MR}} + 3\bar{\nu}^6_{\text{M}} = -4 -6 -8 +18 = 0$ remains.

\item
Band 7 exhibits a second fourfold crossing at $\Gamma$ yielding $\bar{\nu}_{\Gamma}^{7} = -3$. On $\Gamma$--X two Weyl points of identical charge are found, but they appear to be unnecessary for the connectivity, i.e., both bands descending from $\Gamma$ to X emerge from a fourfold degeneracy at $\Gamma$, and thus no band exchanges are needed. In contrast, on $\Gamma$--R the set of crossing points with a total charge of 2 cannot be removed due to substantial gaps surrounding the crossing, $8\nu^7_{\Gamma\text{R}} = 16$. This charge must be compensated by the nodal plane yielding $\nu^7_{\mathrm{npt}} = -( \nu_{\Gamma}^{7} + 8\nu^7_{\Gamma\text{R}} )= -(-3 -8) = 11$.

\end{itemize}
 
The original Chern numbers of the nodal planes $(\nu^1_{\mathrm{npt}},\nu^3_{\mathrm{npt}},\nu^5_{\mathrm{npt}},\nu^7_{\mathrm{npt}}) = (-1,-13,-5,-1)$ have been replaced in the process of applying small, symmetry-preserving perturbations to remove some of the accidental crossing by
$(\bar{\nu}^1_{\mathrm{npt}},\bar{\nu}^3_{\mathrm{npt}},\bar{\nu}^5_{\mathrm{npt}},\bar{\nu}^7_{\mathrm{npt}}) = (1, -3, -3, 11)$. We found that the degeneracy point with $\nu^6_{\Gamma\text{R}} = 2$ is a robust feature affecting the nodal planes in a way which cannot be easily perturbed. Thus, the nodal planes keep non-zero charges not changing monotonically, and one of the four nodal planes at the Fermi energy obtains a higher Chern number than before. 

\subsection{Generic tight-binding models for SG 198}
\label{sec_tight_binding_model}

In this section, we present two generic tight-binding models of SG\,198, one without SOC and one with SOC. To illustrate the arguments presented in the previous sections we calculate the Chern numbers in these generic models. 

To obtain a tight-binding model in agreement with SG~198, we defined a crystal with one 4a Wyckoff position occupied by an s-orbital. Without SOC no internal degrees of freedom are added, otherwise we associate an electron spinor to the orbital. We then defined a tight-binding Hamiltonian by introducing all hopping terms from one site to the others and include complex hopping elements up to next-nearest neighbors, and symmetrized the Hamiltonian by using the action of the space group to add all terms related by symmetry to the hopping terms introduced already. In the resulting matrix only valid terms remained to which we assigned independent but otherwise arbitrary complex values. The resulting band structures are displayed in Fig.~\ref{fig_ModelSG198_Bands}(a) and~\ref{fig_ModelSG198_Bands}(b) without and with SOC, respectively. Using the numerical method presented in Eqs.~\eqref{wilson} to~\eqref{wilson_na} above, we calculated the topological charges of the nodal planes. 

\begin{figure}
\centering
\includegraphics[width = 0.9 \columnwidth]{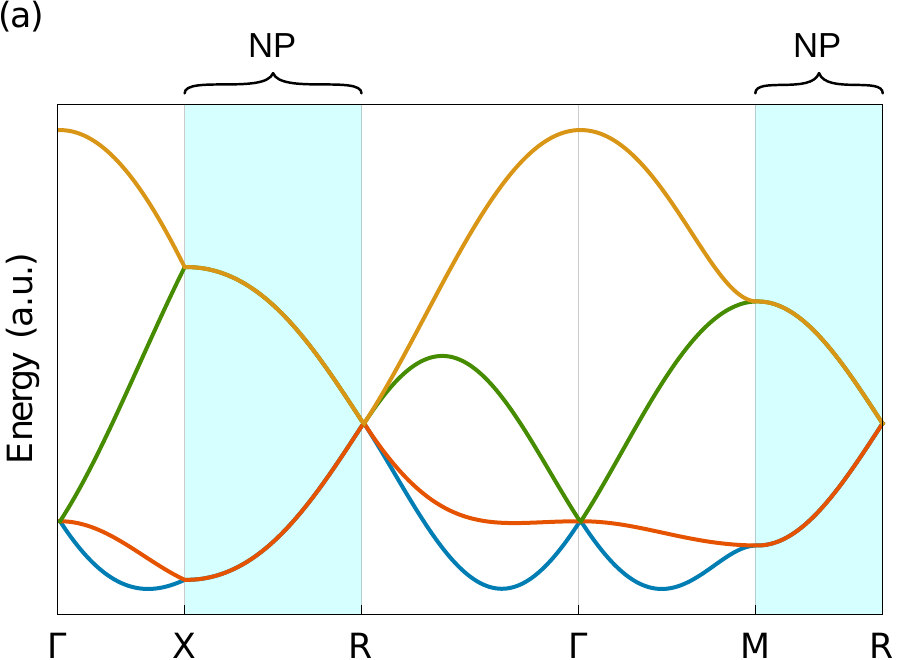} \hspace{.5cm}
\includegraphics[width = 0.9 \columnwidth]{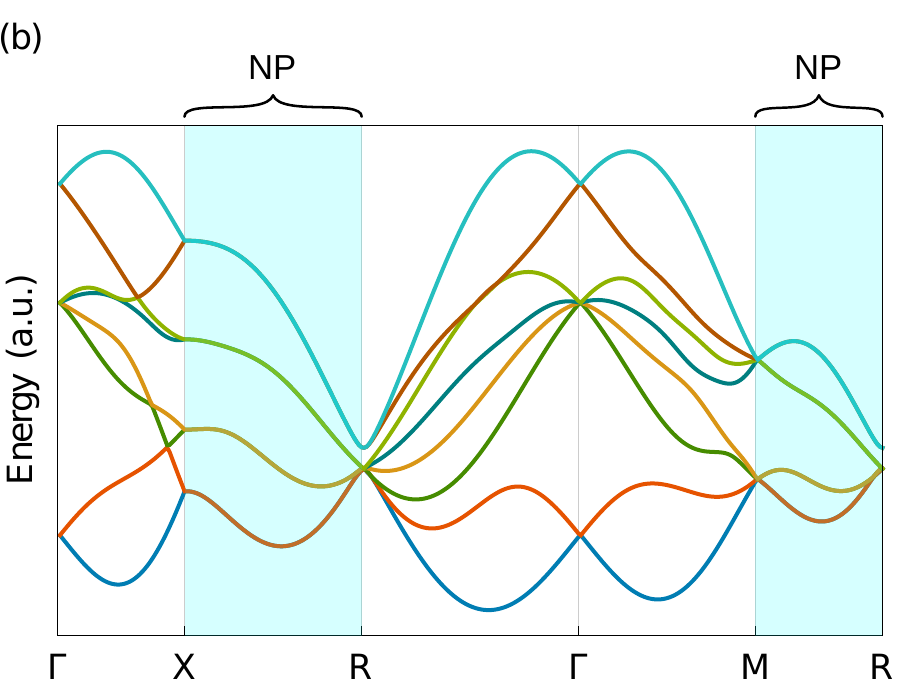}
\caption{Generic tight-binding models of SG\,198 with up to next-nearest neighbor hoppings. (a) Band structure without SOC. (b) Band structure with SOC. }
\label{fig_ModelSG198_Bands}
\end{figure}

Our generic tight-binding model without SOC comprises four bands, which form two nodal planes on the BZ boundary. Unlike for CoSi the model exhibits a threefold and a nondegenerate band at $\Gamma$ only, which is sufficient for a qualitative comparison to CoSi in the vicinity of the Fermi energy. We find the Chern numbers of the trio of nodal planes to be $(\nu^1_{\text{npt,TB}}, \nu^3_{\text{npt,TB}}) = (-4,0)$ where spin degeneracy is included. This corresponds to the expectation; cf Sec.~\ref{sec_CoSI_no_SOC}. Here, the charge of the threefold  degeneracy at $\Gamma$ must be compensated by the nodal plane, where the upper nodal plane is not associated with any charged crossing. 

In our second model we study SG\,198 with a spinful representation, i.e. with SOC, resulting in eight bands, which combine into four nodal planes.  As shown above for CoSi, the $\Gamma$ point exhibits a twofold and a fourfold band crossing. For the generic tight-binding model the accidental crossings do not influence the nodal planes and thus we obtain the Chern numbers $(\nu^1_{\text{npt,TB}}, \nu^3_{\text{npt,TB}}, \nu^5_{\text{npt,TB}}, \nu^7_{\text{npt,TB}}) = (1,3,3,-1)$. The first and fourth nodal planes receive their charges from an uncompensated Weyl point at $\Gamma$, whereas for the second and third nodal plane the fourfold point at $\Gamma$ enforces a charge of 3. 
 
In conclusion, the arguments of Sec.~\ref{sec_CoSI_with_SOC_MinimalPhysicalCrossings} regarding the minimal topological charges of nodal planes in CoSi can be reproduced to a large extent in generic tight-binding models of SG\,198 hence are expected to occur also in other non-magentic representatives of SG\,198, such as RhSi and PdGa. 

\clearpage
\newpage

\section{Experimental data analysis and comparison to literature}
\label{sec:2}

In this section we present information of the experimental methods, the analysis of our experimental data and a comparison of the magnetoresistance oscillations observed in our study with the literature on quantum oscillations in CoSi \cite{Wu_SdH-CPL-2019, Xu_2019_PRB, Wang-dHvA-PRB-2020}. The presentation is organized as follows. We begin with an account of the experimental methods in  Sec.~\ref{sec:methods}, followed in Sec.~\ref{sec:Data-analysis-details} by the magnetoresistance data as recorded and the data treatment to obtain the fast Fourier transform (FFT) spectra shown in the main text. Next, we remind the reader of basic aspects of magnetic breakdown in Sec.~\ref{sec:MB}. In Sec.~\ref{sec:Comparison-QO-studies} we focus on the interpretation of the two oscillation frequencies detected as related to the Fermi surface (FS) sheets at the R-point, denoted as $f_\alpha$ and $f_\beta$ in the main text. In addition, we present data for samples of different quality which suggest that a putative splitting of these frequencies into two pairs of frequencies inferred from an asymmetry reported in Ref.\,\cite{Wang-dHvA-PRB-2020} are not intrinsic and most likely the consequence of reduced sample quality. This is consistent with the analysis of Ref.\,\cite{Xu_2019_PRB}, where the authors rule out any splitting larger than their resolution limit of 2.2\,T. We finally present further implications of the existence of NPs and the relevance of SOC in CoSi.

\subsection{Experimental methods}
\label{sec:methods}

For our study several CoSi single crystals were prepared either using a Te flux as reported in the literature \cite{Wu_SdH-CPL-2019,  Wang-dHvA-PRB-2020}, as well as optical float-zoning under UHV compatible conditions  \cite{Neubauer2011, 2016_Bauer_RevSciInstrum}. To further improve the crystalline quality the ingots and specimen cut from these ingots were annealed under an Ar atmosphere. High sample quality was confirmed in terms of the resistivity, magnetization, specific heat, and x-ray scattering. For the SdH measurements samples were cut with a wire saw. The transport samples exhibited residual resistivity ratios between 16 and 30. The experimental geometry used in our studies is sketched in the inset of Fig.~3\,(a) in the main text. Electrical current was applied along $[110]$. The field was applied in the $[100]$ - $[110]$ plane, perpendicular to the current direction.

The longitudinal and the transverse voltage drops $\rho_{\rm xx}$ and $\rho_{\rm xy}$, respectively, were recorded using a conventional six terminal lock-in technique. Both, $\rho_{\rm xx}$ and $\rho_{\rm xy}$, were evaluated and found to exhibit the same quantum oscillatory signal components. For our studies two cryogenic systems were used. A $^3$He system as combined with a 15\,T superconducting magnet, and a dilution refrigerator equipped with an attocube piezo-driven rotation stage as combined with an 18\,T superconducting magnet system. 

\subsection{Details of the data analysis}
\label{sec:Data-analysis-details}

In this section we report a step-by-step account how the magnetoresistance data was analyzed to determine the frequencies of the Shubnikov-de Haas (SdH) oscillations. The methods used are well established in the analysis of periodic signals.
%
Typical magnetoresistance data are shown in Fig.~\ref{fig:Supp_Data_Analysis}~(a). They are dominated by a monotonically increasing non-oscillatory signal component to which the characteristic SdH oscillations are superimposed. To determine the non-oscillatory signal contribution an $8^{th}$-order polynomial was fitted and subsequently subtracted, where the remaining typical oscillatory part of the signal is shown in in Fig.~\ref{fig:Supp_Data_Analysis}~(b). Because data points were recorded in equidistant steps in $B$ while the oscillations are periodic in $1/B$, the data points were interpolated by a set of equidistant points in $1/B$ prior to the Fourier analysis. The distance between data points was chosen to match the spacing of the data points as recorded at the highest fields. Additionally, zeros were added between 0 and 1~T$^{-1}$ on both sides of the data range recorded experimentally in order to enhance the sampling rate of the fast Fourier transform (FFT) algorithm. 

The interpolated data as a function of $1/B$ are shown in Fig.~\ref{fig:Supp_Data_Analysis}~(c). Note that there is a sharp transition between the boundaries of the data range as recorded down to $1/18~{\rm T}^{-1}$ and the zeros added. These sharp transitions generate additional so-called spectral leakage, resulting in a stronger broadening of the main peaks shown in Fig.~\ref{fig:Supp_Data_Analysis}\,(e) and the possible appearance of side lobes. The spectral leakage can be mitigated through the use of different windowing functions that smooth the sharp transitions at the boundaries of the data range recorded as illustrated in Fig.~\ref{fig:Supp_Data_Analysis}\,(d). For the data shown in the main text and here a Hamming window was used between 1/7 and $1/18~{\rm T}^{-1}$. It exhibits a strong suppression of the first side lobes and is therefore well suited for the analysis of signals comprising two nearby frequencies. As shown in Fig.~\ref{fig:Supp_Data_Analysis}\,(e) the use of a Hamming window significantly reduces spectral leakage while the position and width of the peaks stay the same. Note that the windowing function changes the amplitude. In turn, the same windowing function was applied to all data sets to permit comparison of the amplitude, e.g., for the analysis of the temperature-dependent damping factor $R_T$ described below.

\begin{figure*}[th]
	\includegraphics[width=0.9\textwidth]{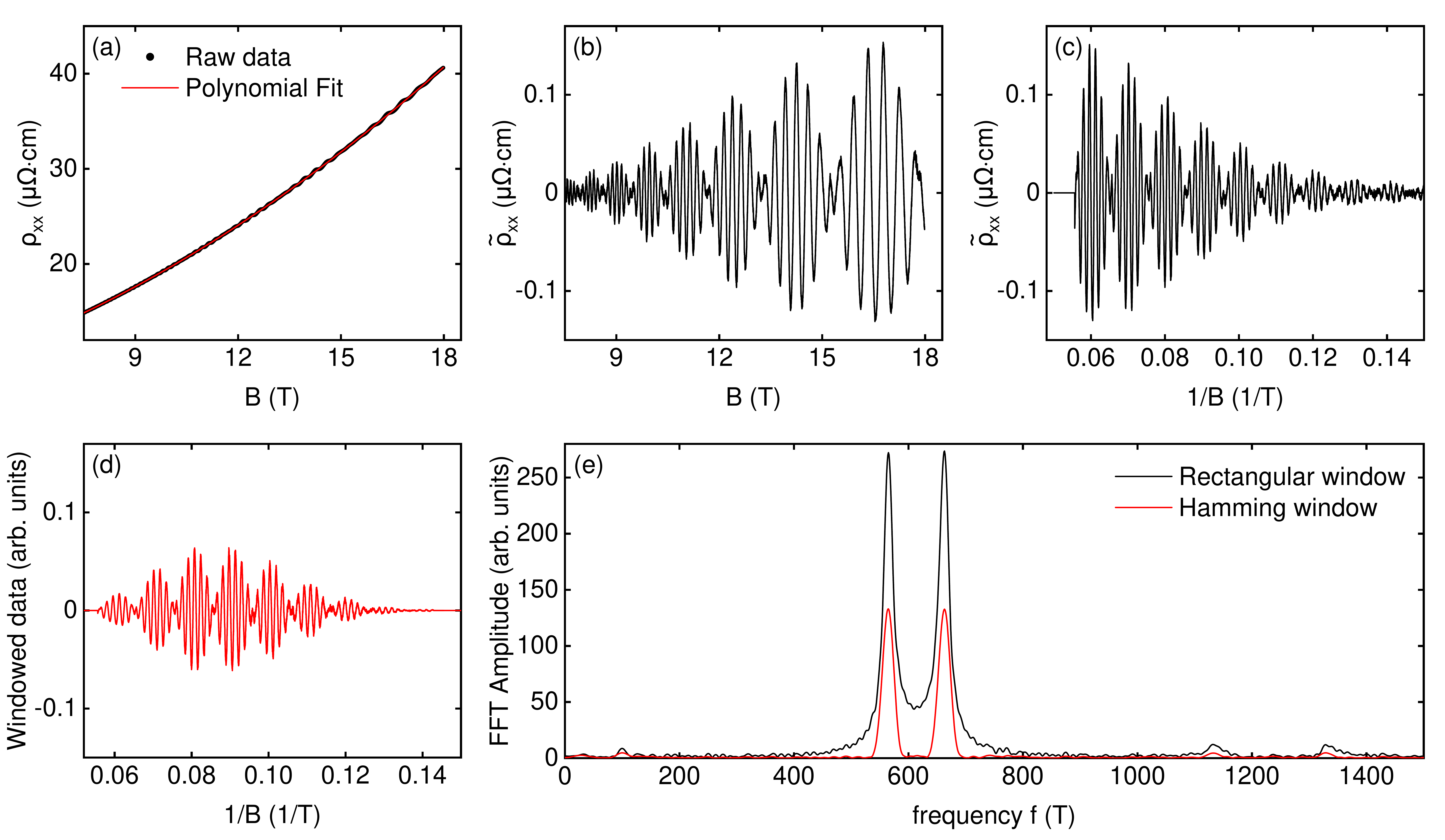}%
	\caption{\label{fig:Supp_Data_Analysis} {\bf Illustration of the analysis of typical experimental data.} (a)~Magnetoresistance as a function of applied magnetic field $B$. A polynomial fit was used to subtract the non-oscillatory signal component. (b)~Oscillatory component of the magnetoresistance as a function of $B$ following subtraction of the non-oscillatory part. (c)~Oscillatory component of the magnetoresistance interpolated on a set of equidistant points as a function of inverse magnetic field. Zeros were added on both sides of the data range. (d)~Data shown in panel (c) following multiplication with a Hamming window between 1/7 and 1/18~T$^{-1}$. (e)~FFT spectrum of the experimental data following application of a rectangular or Hamming window. Note the strong suppression of spectral leakage when using the Hamming window.}
\end{figure*}

The temperature damping factor is given by \cite{Shoenberg1984}
\begin{equation}
R_T=\frac{X}{\sinh(X)} \text{ with } X = \frac{2 \pi^2 p\,m^* k_{\rm B}T}{\hbar\,e\,B},
\label{eq:LK_RT}
\end{equation}
where $p$ is the harmonic number, $m^*$ the cyclotron mass, $k_{\rm B}$ the Boltzmann constant, $T$ the temperature, $\hbar$ the reduced Planck constant, $e$ the electron charge, and $B$ the magnitude of the applied magnetic field.

The cyclotron mass of the extremal orbits was inferred from the temperature dependence of the oscillation amplitudes. In our analysis the amplitudes were determined by means of a Gaussian fit of the FFT peaks inferred from the Hamming-windowed data between $1/7$ and $1/18~{\rm T}^{-1}$. Care was taken when choosing the value of $B$, since the oscillation amplitude changed with $B$ and the applied windowing function alters the amplitudes. For moderately sized windows the mean value of the window in $1/B$ represents a suitable approximation for the calculation of the effective masses. Use of Hamming windows further justified this approximation, as it weighed the data most strongly at exactly this field value being symmetric with respect to it. The estimated uncertainty of the effective masses reported in the main text is 5\%.

\subsection{Magnetic Breakdown}
\label{sec:MB}

Taking into account SOC, the four R-centered FS sheets are close to each other at multiple positions in $k$-space, yielding in principle a multitude of different magnetic breakdown trajectories for most magnetic field directions. Typical magnetic breakdown junctions, where tunneling between adjacent orbits may be expected, are highlighted by colored circles in Fig.~4 of the main text. The probability of switching orbits at a given junction is given by $p=e^{-\frac{B_0}{B}}$, whereas the probability for staying on the initial orbit is $q=1-p$. The breakdown fields $B_0$ for each junction were calculated by means of Chamber's formula \cite{Shoenberg1984}
\begin{equation}
B_0=\frac{\pi \hbar}{2e} \sqrt{\frac{k_g^3}{a+b}} \mbox{ .}
\end{equation}
Here, $k_g$ is the gap in $k$-space, and $a$ and $b$ are the curvatures of the orbits at the junction. Only closed orbits after one cycle are considered here. The resulting angular dispersions as illustrated by the the very faint breakdown branches in Fig.~4\,(d) in the main text were weighted by the probability of their occurrence. In this situation, each pair of orbits has between 0 (for $B\parallel [100]$) and 6 breakdown junctions, depending on the magnetic field direction, resulting in up to $2\times2^6=128$ branches per pair. Most of these orbits are multiply degenerate and have very low weight. Only two almost dispersionless branches carry almost all the spectral weight. They differ in frequency by about 80\,T to 90\,T and have a very low angular dispersion of $\sim 10$\,T .

Experimentally the spectral weight was located at two almost dispersionless frequency branches. For the hypothetical cases shown in Fig.~4\,(c) in the main text, which neglects the effects of the nodal planes, all of the breakdown junctions would be expected between the branches 2 and 3 (light blue and dark red), thus resulting in $0$ to $12$ junctions and consequently up to $2\cdot 12^2=8192$ orbits. Due to this large number, the associated frequencies are not displayed explicitly in Fig.~4\,(c) in the main text. However, all of these breakdown branches are expected in the shaded area between branches 2 and 3.

As discussed in the main text, only the scenario taken into account SOC, the existence of nodal planes, and the occurrence of magnetic breakdown between adjacent FS sheets is consistent with the experimental data.

\subsection{Comparison with the literature}
\label{sec:Comparison-QO-studies}

In our study we focused on the R-centered FS sheets of CoSi in order to investigate the existence and relevance of topological NPs in SG 198. As our main result we found that quantum oscillations arising from extremal orbits on the R-centered FS pockets provide clear evidence of, both, SOC-induced band splitting and the existence of topological nodal planes. In comparison, the interpretation of related experimental data reported in the literature did not consider the presence of nodal planes, and either ignored or misinterpreted the effect of SOC on the quantum oscillation spectra. In the following, we address these differences of the interpretation in detail.

In studies of quantum oscillations in CoSi reported in the literature, probing the magneto-thermopower and Nernst voltage \cite{Xu_2019_PRB} and magnetoresistance \cite{Wu_SdH-CPL-2019}, the detection of two main frequencies was incorrectly attributed to two spin-degenerate FS sheets centered around the R-point. This corresponds to a scenario neglecting both SOC and nodal planes. However, closer inspection reveals that the actual prediction for this scenario shown in Fig.~4\,(a) in the main text is not correct and any partial qualitative agreement fortuitous. 
As described in the main text, for a technically correct calculation the experimental observation of two dominant, nearly dispersionless oscillation frequencies is \textit{not} consistent with two spin-degenerate FS sheets centered around the R-point. Rather, it agrees very well with a SOC-induced splitting of the bands around the R-point taking topological nodal planes into account.

This can be illustrated, e.g., for the case where $B$ is parallel to $[100]$. Here, the main extremal orbits are located in the $k_x=\pi$ plane which is nodal, i.e., the bands are (at least) twofold degenerate. Thus, when not taking into account SOC the four spin-degenerate bands crossing the FS form a single fourfold degenerate band on the Brillouin zone (BZ) boundary that gives rise to a single oscillation frequency. In addition a small branch corresponding an extremal orbit running in a "ridge" of one FS sheet exists only very close to the $[100]$ direction. This branch is very close in frequency (with a difference $<10$\,T) to the degenerate main orbits [Fig.~4\,(a)]. In comparison, when including SOC and the nodal planes, the four bands split into two twofold degenerate bands resulting in two oscillation frequencies spaced about 80\,T apart. This behavior is observed experimentally.

Extremal orbits perpendicular to directions other than $[100]$ are not twofold degenerate everywhere. However, they still have to cross at the NPs at the BZ boundary. At these crossings the correct band connectivity must be taken into account. As explained in the main text and Ref.~\cite{MnSi_nodal_plane}, the underlying wave-functions of two bands crossing on a NP are orthogonal and therefore transitions between them are suppressed. Without SOC, this would result in two extremal orbits enclosing the same area in momentum space, as depicted in Figs. 2 and 4 in the main text, and thus to a \textit{single} frequency branch for the entire angular range between [100] and [110] [Fig.~4\,(b)]. Therefore, the detection of two almost dispersionless frequency branches related to the FS around the R-point is a clear signature of four FS sheets and significant SOC-induced splitting of the bands in CoSi.


\begin{figure*}
	\includegraphics[width=0.6\textwidth]{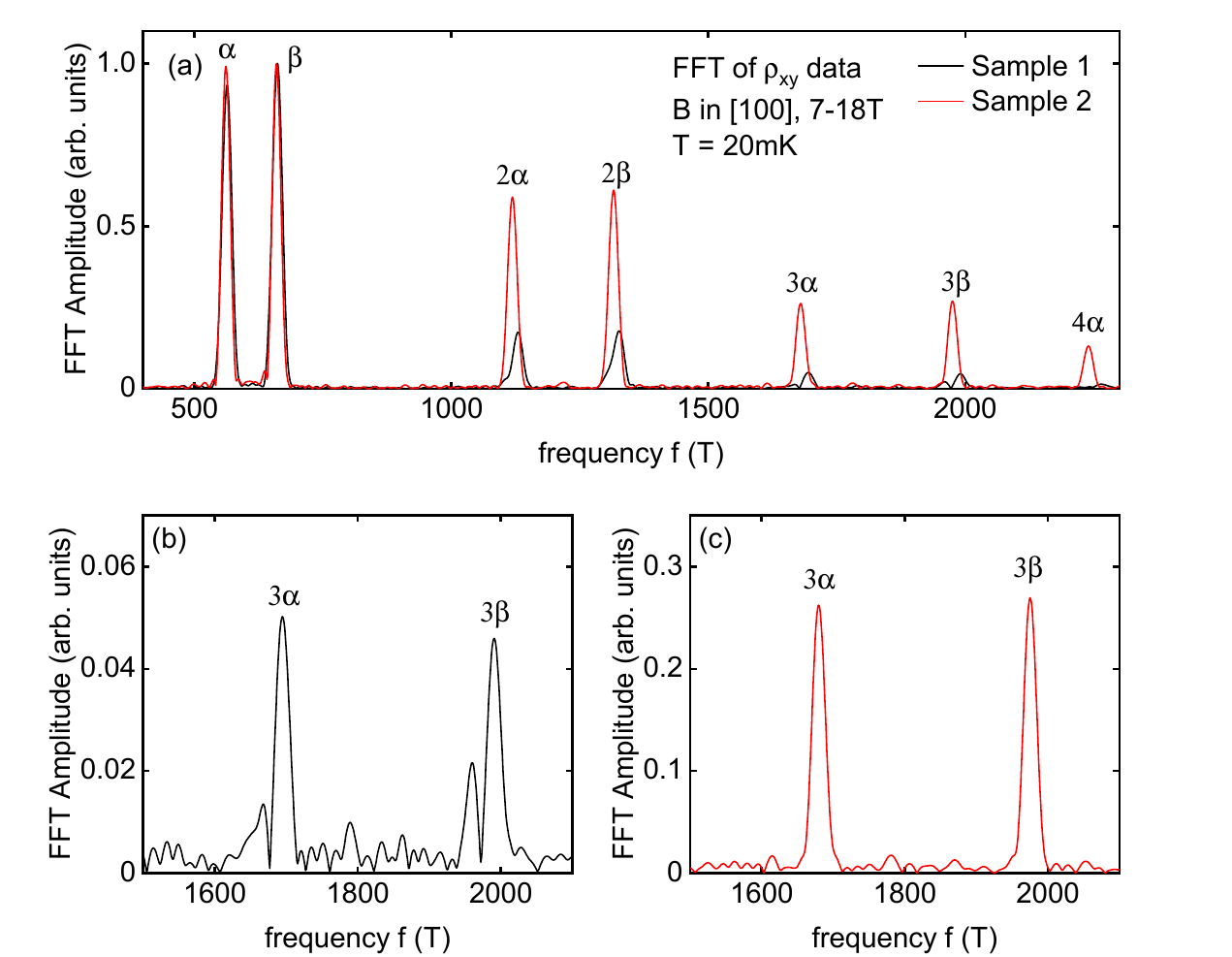}%
	\caption{\label{fig:Asymmetry_Analysis} 
	{\bf Analysis of a possible asymmetry of the FFT peaks.} (a)~Typical FFT spectra of two different samples with different RRR, where sample 2 exhibits the higher RRR. (b)~Detailed view of the third order harmonics of sample 1. A clear splitting of the frequencies is visible. (c)~Detailed view of the third order harmonics of sample 2. No splitting visible suggesting that the asymmetry is not due to an intrinsic splitting of oscillation frequencies.}
\end{figure*}

A recent study of de~Haas-van~Alphen (dHvA) oscillations in CoSi~\cite{Wang-dHvA-PRB-2020} reported the observation of a SOC-induced band splitting. However, it is important to note that the authors did not take into account the existence of NPs and the associated band connectivities at the NPs. Moreover, unfortunately the authors did not calculate the dispersion of the frequency branches to be expected in the scenario they proposed. As reported in Fig.~4\,(c) in the main text, this dispersion would be rather strong and different for different branches and thus inconsistent with experiment. 

The interpretation proposed in Ref.\,\cite{Wang-dHvA-PRB-2020} is based on the detection of asymmetrical peaks, which is attributed to a splitting of the two main frequencies into a total of four frequencies with each asymmetrical peak reflecting two distinct frequencies separated by $\sim$10\,T. This corresponds to a SOC splitting of only a few meV. This contrasts the strength of the SOC of 25 to 35\,meV in the bulk inferred from STM data \cite{2019-Yuan-Science-Advances}, as well as the value of $\sim 15\,\text{meV}$ observed in our calculations.

To explore the putative existence of such a tiny splitting of the oscillation frequencies as an intrinsic property, we analyzed SdH spectra we recorded in several samples exhibiting different residual resistivity ratios (RRRs). Shown in Fig.~\ref{fig:Asymmetry_Analysis} are typical FFT spectra inferred from the Hall signal of two samples of different quality, where the magnetoresistance shows a similar behavior albeit the higher harmonics are not as well developed for the sample with lower RRR. A Hamming window was used between 1/7 and 1/18~T$^{-1}$ to reduce spectral leakage. 

For both samples the higher harmonics up to the third order may be clearly resolved as shown in Fig.~\ref{fig:Asymmetry_Analysis}~(b) and \ref{fig:Asymmetry_Analysis}~(c). The amplitude of the higher harmonics of sample 2 decreases less strongly with increasing harmonic number reflecting the higher crystalline quality of this sample. If there would be an asymmetry in the main peaks $\alpha$ and $\beta$ originating from two distinct quantum oscillation frequencies for each peak, it should be observable more clearly in the higher harmonics, because the separation of the two peaks scales linearly with the harmonic number while the width of the peaks for the main frequencies and its harmonics is nearly the same. 

Analyzing our data for different angles of the applied magnetic field (not shown here) we consistently found a splitting of the peaks in sample 1 and no splitting in sample 2. We also note that in Ref.~\cite{Xu_2019_PRB} the authors explicitly looked for a splitting of the main frequencies and concluded that no splitting larger than 2.2~T is present in their samples. In summary, while the origin of two close oscillation frequencies around $\alpha$ and $\beta$ in samples of lower RRR deserves further investigation, we consider it not to be an intrinsic property of the electronic structure of single crystal CoSi since it is not detectable in the sample exhibiting the highest RRR.


\begin{figure*}[t]
	\includegraphics[width=0.6\textwidth]{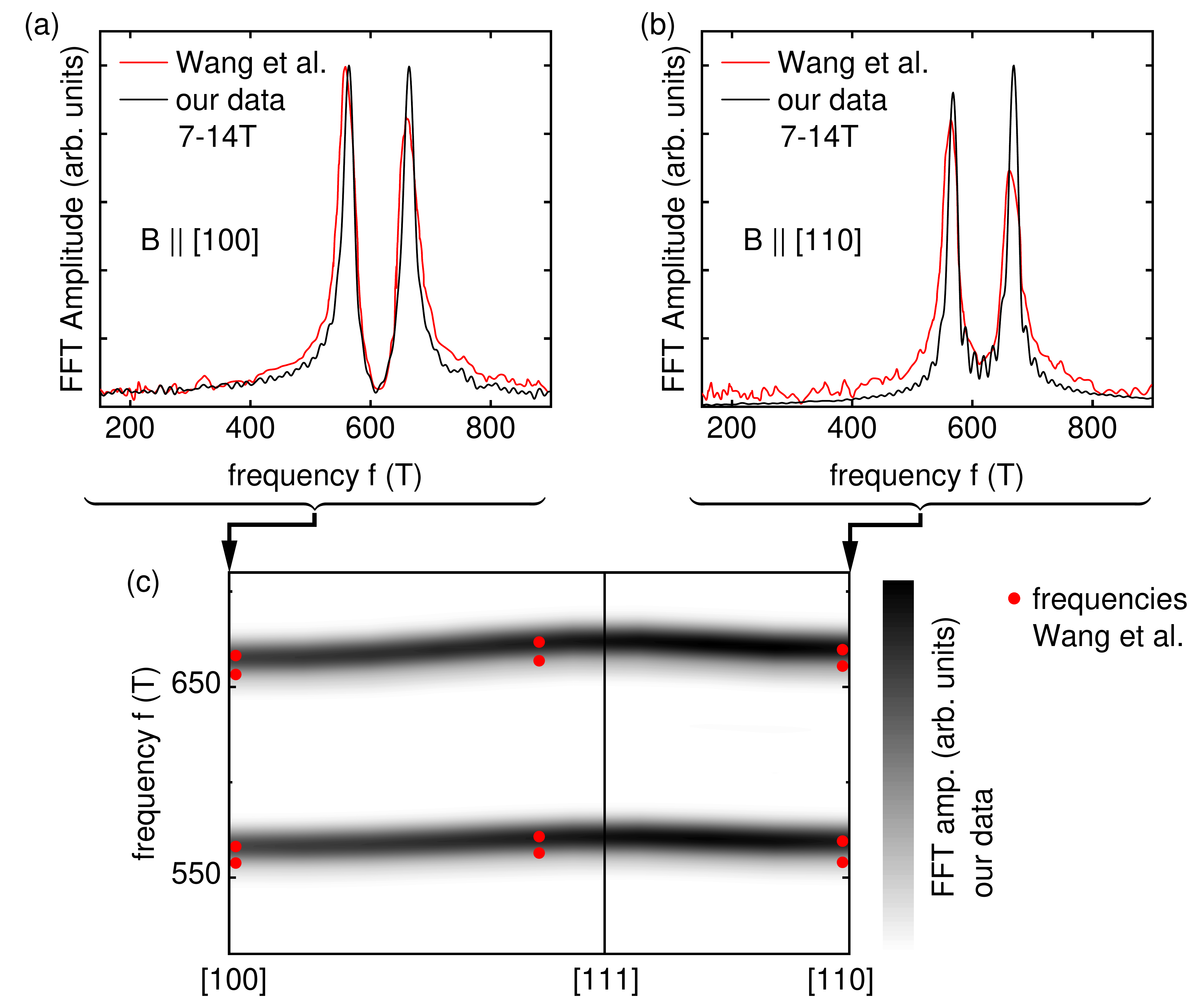}%
	\caption{\label{fig:Supp_WangvsHuber} {\bf Comparison of our FFT data with the data of Wang \emph{et al.} (Ref.~\cite{Wang-dHvA-PRB-2020}).} (a) and (b)~ FFT data of Wang \emph{et al.} (red) and our data (black) measured on sample 1 and analyzed in a rectangular window ranging from $1/7$ to $1/14$~T$^{-1}$ with $B$ along [100] and [110], respectively. (b) Angular dispersion of the frequency branches associated with the R-centered FS pockets. The FFT-peak positions (symbols) given in Ref.~\cite{Wang-dHvA-PRB-2020} are consistent with our data (grayscale).}
\end{figure*}

While the sample dependence of the putative asymmetry of the FFT peaks provides strong empirical evidence against an intrinsic frequency splitting due to SOC, it is nonetheless instructive to consider the possible implications for the experimental behavior, if there would be such a small SOC-induced frequency splitting of the frequency branches $\alpha$ and $\beta$ in which the band connectivity at the NPs is neglected. 

We begin by noting that the frequencies presented in Ref.~\cite{Wang-dHvA-PRB-2020} are roughly consistent with the frequencies we observed in our measurements. This is highlighted in Fig.~\ref{fig:Supp_WangvsHuber}\,(a) and (b) where we reproduce the FFTs of Wang \emph{et al.} for $B\parallel$ [100] and $B\parallel$ [110] and the FFTs performed with the same rectangular window function Wang \emph{et al.} used on our data as recorded in sample 1. Here the corresponding peaks reported by Wang \emph{et al.}  and in our data coincide. Moreover, as shown in Fig.~\ref{fig:Supp_WangvsHuber}\,(c) for the selected field orientations reported by Wang \emph{et al.} the frequencies are in excellent agreement with the angular dispersion we observed, highlighting the overall agreement.

It follows that the arguments given in the main text in Fig.~4 apply equally well for the data of Wang \emph{et al.}. Focusing on the case of extremal orbits perpendicular to [110] depicted in Fig.~4\,(c) in the main text and ignoring the correct band connectivity at the NPs, the DFT calculation predicts two pairs of extremal orbits leading to four distinct frequency branches away from the [100]. Close to [100], these branches would merge pairwise, while at the same time two additional branches would arise from the maximal orbits running on the ridges of the constrictions of the second and fourth FS sheet.

The frequency difference between the minimal and maximal orbits close to [100] is indeed ~8T as calculated in Ref.~\cite{Wang-dHvA-PRB-2020}. However, the minimal and maximal orbits on each sheet quickly merge into one frequency when rotating away from [100], i.e., when the orbit is not running inside a ridge anymore. At the same time, the two-fold degeneracy of the R-centered orbits would be lifted, leading to a branch splitting of 60~T to 85~T over most of the angular range. In contrast, the separation of frequencies forming the asymmetrical peaks is about 10~T over the whole angular range in our sample exhibiting the frequency splitting as well as in the data reported in~\cite{Wang-dHvA-PRB-2020}. That is, the measured frequency splitting, if present at all, is almost an order of magnitude smaller than the splitting expected in the scenario proposed in Ref.~\cite{Wang-dHvA-PRB-2020}, showing that this scenario is inconsistent with the data.

We finally note that the extracted effective masses of the two main frequencies reported in~\cite{Wang-dHvA-PRB-2020} differ by a factor of almost 3 from the values obtained in~\cite{Xu_2019_PRB,Wu_SdH-CPL-2019} and the consistent values found from our data.

To conclude, the detection of two dominant oscillation frequencies related to the FS sheets centered at the R-point in CoSi is a clear signature of SOC-induced band splitting by $\sim15\,{\rm meV}$ and the presence of four FS sheets when the correct band connectivity at the NPs is taken into account. As emphasized above, the small splitting of the two dominant frequencies observed in CoSi samples with lower RRR deserves further investigation, but does not appear to be an intrinsic property of the Fermi surface of CoSi.


%